\documentclass[11pt,a4paper]{article}
\usepackage{amsmath,amsfonts,epsfig}
\usepackage{amssymb}
\usepackage{mathrsfs}
\usepackage{hyperref}
\begin{document}
\numberwithin{equation}{section}
\setlength{\unitlength}{.8mm}

\begin{titlepage} 
%\begin{flushright}
%version1\\
%\today
%\end{flushright}
\vspace*{0.5cm}
\begin{center}
{\Large\bf Norm of Bethe-wave functions in the continuum limit}
\end{center}
\vspace{1.5cm}
\begin{center}
{\large \'Arp\'ad Heged\H us}
\end{center}
\bigskip

\vspace{0.1cm}

\begin{center}
Wigner Research Centre for Physics,\\
H-1525 Budapest 114, P.O.B. 49, Hungary\\ 
\end{center}
\vspace{1.5cm}
\begin{abstract}
The 6-vertex model with appropriately chosen alternating inhomogeneities gives 
the so-called light-cone lattice regularization of the sine-Gordon (Massive-Thirring) 
model. In this integrable lattice model we consider pure hole states above the 
antiferromagnetic vacuum and express the norm of Bethe-wave functions in terms of the hole's positions 
and the counting-function of the state under consideration. 
In the light-cone regularized picture pure hole states correspond to pure soliton (fermion) states of the 
sine-Gordon (massive Thirring) model.
Hence, we analyze the continuum limit of our new formula for the norm of the Bethe-wave functions. 
We show, that the physically most relevant determinant part of our formula can be expanded in the large volume limit 
and turns out to be proportional to the Gaudin-determinant of pure soliton states in the sine-Gordon model 
defined in finite volume.
\end{abstract}

\end{titlepage}

%%%%%%%%%%%%%%%%%%%%%%%%%%%%%%%%%%%%%%%%%%%%%%%%%%%%%%%%%%%%%%
\section{Introduction} \label{intro}

Recently, finite volume form-factors in integrable quantum field theories attract interest because of their 
relevance in the solution of the planar AdS/CFT correspondence \cite{BJsftv,Bjhhl}.

There are two basic approaches to finite volume form-factors in integrable quantum field theories. 
The first approach initiated in \cite{PT08a,PT08b} describes the finite volume form-factors in the form of a large 
volume series built from the infinite volume form-factors \cite{Smirnov} of the theory. 
The second one uses an integrable lattice regularization 
of the quantum field theory under consideration and provides with exact expressions for the finite volume form-factors.

The first approach proved to be 
particularly successful in purely elastic scattering theories. In this class of theories an all order large volume 
series was proposed for the diagonal form-factors \cite{LM99,Pozsg13,PST14}, and the first order exponentially small in volume 
corrections of the non-diagonal form-factors have been determined \cite{Pmu,BBCL}. 
However, up to now in non-diagonally scattering theories 
 the large volume series based method made it possible to determine finite volume form-factors 
only up to polynomial corrections in the inverse of the volume \cite{FT, FPT11,Palmai13}. 

The second approach based on an integrable 
lattice regularization of the quantum field theory under consideration, 
led to remarkable results in the sine-Gordon (Massive-Thirring) model. In this framework finite volume 1-point functions \cite{BS4,BS6},  
ratios of infinite volume form-factors \cite{BS7} and the diagonal finite volume solitonic matrix elements of the $U(1)$ 
current \cite{En} and the trace of the stress-energy tensor \cite{En1} have been determined. 
In \cite{En,En1} the exact results served by the lattice  were rephrased as a large volume series admitting a similar mathematical structure to those 
arising in the purely elastic scattering theories \cite{Pozsg13,PST14}. 

In the lattice computations, ratios of determinants of square matrices with size being comparable to the number of lattice points, should be computed. 
Fortunately, in some special cases, mostly when diagonal form-factors are considered, 
the computation of these ratios of determinants simplify to mathematically treatable problems. This makes  
 the determination of diagonal form factors possible \cite{En,En1}, because in this case the ratio of huge 
determinants reduces to 
the determinant of a low dimensional matrix 
with entries given by the solutions of certain linear integral equations with kernels containing the counting-function of the model. 

However, these helpful simplifications are not present, when non-diagonal form-factors are considered. In this case one cannot avoid to determine 
the determinants of square matrices with size being comparable to the number of lattice points. 

In this paper we consider the simplest determinant arising in the light-cone lattice regulrization \cite{ddvlc} based computation of non-diagonal form-factors 
in the sine-Gordon (massive Thirring) model. This is the so-called Gaudin-determinant 
$\text{det} \Phi$, which determines the norm square of a Bethe-eigenstate 
$|\vec{\lambda}\rangle$ (\ref{sajatvec}) %$\quad B(\lambda_1)...B(\lambda_m)|0\rangle,$ 
through the formula \cite{Gaudin0,Gaudin1,Korepin}:
\begin{equation} \label{norm0}
\begin{split}
||\vec{\lambda}||^2=\langle \vec{\lambda}|\vec{\lambda} \rangle=v_0 \, \text{det} \Phi, \qquad 
\end{split}
\end{equation}
where the Gaudin-matrix $\Phi,$ and the prefactor $v_0$ are given by formulas (\ref{Phiab}) and (\ref{v0}), respectively.

This determinant representation plays an important role in the determination of non-diagonal form-factors, since in this case 
matrix elements of local operators between normalized eigenstates of the Hamiltonian are considered.
For a local operator $\hat{O},$ a non-diagonal form-factor is given by the formula:
\begin{equation} \label{FO1}
\begin{split}
{\mathcal F}_{O}(\vec{\mu},\vec{\lambda})=\frac{\langle\vec{\mu}|\hat{O}|\vec{\lambda}\rangle}{||\vec{\mu}|| \! \cdot \! ||\vec{\lambda}||},
\end{split}
\end{equation}
where $|\vec{\mu}\rangle$ and $|\vec{\lambda}\rangle$ are Bethe-eigenvectors.
Here, the appearance of the norms of the sandwiching Bethe-eigenstates is the consequence of the fact that in general 
the Bethe-eigenstates are not normalized to one.

In this paper we compute $||\vec{\lambda}||^2$ for pure hole states by expressing both $v_0$ in (\ref{v0}) and ${\det} \Phi$ in terms of the 
positions of holes and of the counting-function of the model. We also investigate the continuum limit of the Gaudin-determinant ${\det} \Phi$ and show that 
its most complicated determinant part is proportional to a product of two determinants. 
The first one is the finite-dimensional Gaudin-determinant of the soliton states described by the holes and the second one is a functional determinant,
which can be expanded in the large volume limit in a straightforward manner.

The paper is organized as follows. In section 2. a brief summary of the quantum inverse scattering method 
can be found. In section 3. the nonlinear integral equations satisfied by the counting-function are 
summarized. In section 4. the dressed Gaudin-matrix of the sine-Gordon solitons are described. 
Section 5. contains some summation formulas being necessary for the computation of the norm of Bethe-eigenstates. 
Sections 6. and 7. contain the computation of the Gaudin-determinant on the lattice and in the continuum limit, 
respectively. 
Section 8. contains the computation of the multiplicative factor $v_0$ entering the Gaudin-formula. 
We close the body of our paper with the summary of our results. 
The  paper also contains two appendices, in which some formulas being helpful for the computations 
of the paper are collected.

\section{Quantum inverse scattering description} \label{QISMsect}

We consider the 6-vertex model defined by the $R$-matrix:
\begin{equation} \label{Rmatrix}
 R(\lambda)=\begin{pmatrix} 1 & 0 & 0 & 0 \\
 0 & \tfrac{\sinh(\lambda)}{\sinh(\lambda-i \gamma)} & \tfrac{\sinh(-i \gamma)}{\sinh(\lambda-i \gamma)} & 0 \\
0 & \tfrac{\sinh(-i \gamma)}{\sinh(\lambda-i \gamma)}  & \tfrac{\sinh(\lambda)}{\sinh(\lambda-i \gamma)} & 0 \\
0 & 0 & 0 & 1 
\end{pmatrix}, \qquad 0<\gamma< \pi.
\end{equation}
The monodromy-matrix of the model is given in terms of the $R$-matrix in the usual way: %\cite{KMT00}: 
\begin{equation} \label{monodromy}
 T(\lambda|\vec{\xi})=R_{01}(\lambda-\xi_1)\, R_{02}(\lambda-\xi_2)\, ...R_{0N}(\lambda-\xi_N)=\begin{pmatrix} A(\lambda) & B(\lambda) \\
 C(\lambda) & D(\lambda) \end{pmatrix}_{[0]},
 \end{equation}
where $\xi_n$s are the inhomogeneities of the model. In this paper we choose inhomogeneities with alternating real parts:
\begin{equation}\label{xis}
 \xi_n=\rho_n -i \tfrac{\gamma}{2}, \qquad \rho_n=(-1)^n \rho_0, \qquad n=1,..,N, \qquad \rho_0\in {\mathbb R}.
 \end{equation}
 The homogeneous case $\rho_0=0,$ correspond to the XXZ spin-chain and the inhomogeneous case with appropriate choice for $\rho_0$ 
 describes a lattice regularization for the Massive-Thirring (sine-Gordon) model \cite{ddvlc,deVegaIJMP}.
The monodromy matrix acts as a $2 \times 2$ matrix on the auxiliary space: $V_0\sim {\mathbb C}^2$ such that its entries
act on the quantum space of the model ${\cal H}=\otimes_{i=1}^N {\mathbb C}^2.$ 
The transfer matrix of the model is defined by the trace of the monodromy matrix over the auxiliary space $V_0$:
\begin{equation} \label{trans}
{\cal T}(\lambda|\vec{\xi})=\text{Tr}_0 \, T(\lambda|\vec{\xi}). %\qquad 
\end{equation}
The transfer matrices form a commutative family of operators, since they commute at different values of the spectral parameter $\lambda:$
\begin{equation} \label{comm1}
\left[ {\cal T}(\lambda|\vec{\xi}),{\cal T}(\lambda'|\vec{\xi}) \right]=0, \qquad \forall \lambda, \lambda' \in 
{\mathbb C}.
\end{equation}
This property ensures the integrability of the model and makes it possible to find the eigenvalues and eigenstates of the 
transfer-matrix in a purely algebraic way, with the help of the  
Algebraic Bethe-Ansatz method \cite{FST79}. In this method the eigenstates of the transfer matrix 
(\ref{trans}) are  constructed in the form:
\begin{equation} \label{sajatvec}
|\vec{\lambda}\rangle=|\lambda_1,\lambda_2,..,\lambda_m \rangle=B(\lambda_1)\, B(\lambda_2)\, ...B(\lambda_m)\,|0 \rangle, \qquad  S_z|\vec{\lambda}\rangle=(\tfrac{N}{2}-m)|\vec{\lambda}\rangle,
\end{equation}
where $|0\rangle$ denotes the completely ferromagnetic state of the model with all spins up, 
the operator $B$ denote the $12$-element of the monodromy matrix (\ref{monodromy}),  and the the parameters $\lambda_j$
satisfy the Bethe-equations as follows:
\begin{equation} \label{BAE}
\prod\limits_{i=1}^N \, \frac{\sinh(\lambda_a-\xi_i-i \gamma)}{\sinh(\lambda_a-\xi_i)} \, 
\prod\limits_{b=1}^m \, \frac{\sinh(\lambda_a-\lambda_b+i \gamma)}{\sinh(\lambda_a-\lambda_b-i \gamma)}=-1, \qquad a=1,...,m.
\end{equation}
Here $N$ denotes the number of lattice points and $m$ stands for the number of Bethe-roots.

In the Algebraic Bethe Ansatz approach the physical quantities of the model can be expressed in terms of the Bethe-roots. 
In this paper we are interested in the norm of a Bethe-eigenstate (\ref{sajatvec}). This is given by the famous Gaudin-formula \cite{Gaudin0,Gaudin1,Korepin}:
\begin{equation} \label{norm1}
\begin{split}
||\vec{\lambda}||^2=\langle \vec{\lambda}|\vec{\lambda} \rangle=v_0 \, \text{det} \Phi, \qquad \text{with} \quad 
|\vec{\lambda}\rangle=\quad B(\lambda_1)...B(\lambda_m)|0\rangle,
\end{split}
\end{equation}
where the prefactor $v_0$ and the matrix of the Gaudin-determinant are given by the formulas:
\begin{equation} \label{v0}
\begin{split}
v_0=\frac{\prod\limits_{j,k=1}^m \, \sinh(\lambda_j-\lambda_k-i \, \gamma)}{\prod\limits_{j>k}^m \sinh(\lambda_j-\lambda_k) \, \sinh(\lambda_k-\lambda_j)},
\end{split}
\end{equation}
%{\prod\limits_{k \neq j}^m \, \sinh(\lambda_j-\lambda_k)},
\begin{equation} \label{Phiab}
\begin{split}
\Phi_{ab}=-i \, \partial_{\lambda_a} Z(\lambda_b), \qquad a,b=1,...m.
\end{split}
\end{equation}
In (\ref{Phiab}) $Z(\lambda)$ denotes the counting-function corresponding to the state $|\lambda\rangle.$ 
For the set of inhomogeneities (\ref{xis}), it is defined by the formula \cite{ddv97}:
\begin{equation} \label{countfv}
Z(\lambda)=\frac{N}{2} \, \left( \phi_1(\lambda-\rho_0)+\phi_1(\lambda+\rho_0)\right)-\sum\limits_{k=1}^{m} \, \phi_2(\lambda-\lambda_k),
\end{equation}
where $\phi_\nu(\lambda)$ is given in its fundamental domain $|\text{Im} \, \lambda| < \nu$ by the formula:
\begin{equation} \label{phinu}
\phi_\nu(\lambda)=-i \, \ln \frac{\sinh(i \tfrac{\gamma}{2} \nu -\lambda)}{\sinh(i \tfrac{\gamma}{2} \nu +\lambda)}, \qquad 0<\nu, \qquad \phi_{\nu}(0)=0, \qquad |\text{Im} \, \lambda| < \nu.
\end{equation}
This function can be continued analytically to the whole complex plane by requiring $\Phi_\nu(\lambda)$ to be an odd function with all discontinuities running parallel to the 
real axis \cite{ddv97}. With this definition of $\Phi_\nu(\lambda),$ formula (\ref{countfv}) admits a definition of $Z(\lambda)$ being valid on the 
whole complex plane. This definition allows one to reformulate the Bethe-equations (\ref{BAE}) in their logarithmic form:
%%%%%%%%%%%%%%%%%%%%%%%%%%%%%%%%%%%
\begin{equation} \label{BAEZ}
Z(\lambda_a)=2 \pi \, I_a, \quad  \quad I_a\in \mathbb{Z}+\tfrac{1+\delta}{2}, \qquad \delta=m \,\, (\text{mod} \, 2),
\qquad a=1,..,m.
\end{equation}
In this formulation, to each Bethe-root $\lambda_a$ an integer or half-integer quantum number $I_a$ can be assigned. 
The actual value of the parameter $\delta \in \{0,1\}$ makes difference between the different types of quantizations. 
In this paper we consider pure hole states states above the antiferromagnetic vacuum. These states contain only 
real Bethe-roots.
%such that the quantum numbers of the Bethe-roots fill completely the whole allowed range 
%$\left[Z_{\lambda}(-\infty)/ 2 \pi,Z_{\lambda}(\infty)/ 2 \pi \right]$ in the real axis.

The antiferromagnetic-vacuum of the model is a $\delta=0$ state 
with half-integer quantum numbers filling completely the whole allowed range 
$\left[Z(-\infty)/ 2 \pi,Z(\infty)/ 2 \pi \right].$  %in the real axis.
The excitations above this sea of real roots are characterized by complex Bethe-roots and holes. 
The holes are such special real solutions of
 (\ref{BAEZ}), which are not Bethe-roots{\footnote{Namely, they do not enter in the definition of
 $Z(\lambda)$ in (\ref{countfv}).}}, thus they satisfy the quantization equations as follows: 
\begin{equation} \label{BAEH}
Z({h}_k)=2 \pi \, I_k, \quad  \quad I_k\in \mathbb{Z}+\tfrac{1+\delta}{2}, \qquad k=1,..,m_H,
\end{equation} 
where $h_k$ denotes the positions of the holes and their number is denoted by $m_H.$
In the continuum limit these hole excitations describe the fermions and the solitons of 
the Massive-Thirring and of the sine-Gordon models, respectively. 

Using the definition of (\ref{countfv}) for the counting-function, the entries of the Gaudin-matrix (\ref{Phiab}) 
can be given explicitly by the following formula:
\begin{equation} \label{Gm1}
\begin{split}
\Phi_{ab}=-i \, \left(Z'(\lambda_b) \, \delta_{ab}+ 2 \pi \, K(\lambda_a-\lambda_b)\right), \qquad a,b=1,...,m,
\end{split}
\end{equation}
where 
\begin{equation} \label{K1}
K(\lambda)=\frac{1}{2 \pi} \frac{\sin(2 \, \gamma)}{\sinh(\lambda-i \, \gamma)\, \sinh(\lambda+i \, \gamma)},
\end{equation}
and $\delta_{ab}$ stands for the Kronecker-delta symbol. 

As we already mentioned, the appropriate choice of the inhomogeneity parameter $\rho_0$ in (\ref{xis}) makes it possible to describe 
the sine-Gordon or Massive-Thirring models. These continuum quantum field theories are defined by the Lagrangians:
\begin{equation}
\label{sG_Lagrangian}
{\cal L}_{SG}= \displaystyle\frac{1}{2}\partial _{\nu }\Phi \partial ^{\nu }\Phi +\displaystyle \alpha_0  \left( \cos \left( \beta \Phi \right)-1 \right), \,  \qquad 0<\beta^2<8 \pi,
\end{equation}
\begin{equation}
\label{mTh_Lagrangian}
\!\!\!\!\!\!\!\!\!\!\!\!\!\!\!\!\! \!\!\!\!\!\!\!\!\!\!\!\!\!\!\!\!\! \!\!\!\!\!\!\!
{\cal L}_{MT}= \bar{\Psi }(i\gamma _{\nu }\partial ^{\nu }-m_{0})\Psi -\displaystyle\frac{g}{2}\bar{\Psi }\gamma^{\nu }\Psi \bar{\Psi }\gamma _{\nu }\Psi \,,
\end{equation}
where $m_0$ and $g$ denote the bare mass and the coupling constant of the theory, respectively.
As usual, $\gamma_\mu$s stand for the $\gamma$-matrices satisfying  
the algebraic relations: $\{\gamma^\mu,\gamma^\nu\}=2 \eta^{\mu \nu}$ with $\eta^{\mu \nu}=\text{diag}(1,-1)$.
These two quantum field theories are identical in their even topological charge sector \cite{s-coleman,klassme} if their
coupling constants are related by the formula:
\begin{equation} \label{gbeta}
1+\frac{g}{4 \pi}=\frac{4 \pi}{\beta^2}.
\end{equation}
In \cite{ddvlc} it has been shown, that the even topological charge sector of the Massive-Thirring model can be described as the continuum limit of the 6-vertex model 
with alternating inhomogeneities, provided the inhomogeneity parameter $\rho_0$ in (\ref{xis}) 
is given by the formula:
\begin{equation} \label{rho}
\rho_0=\tfrac{\gamma}{\pi} \, \ln \tfrac{4}{{\cal M} \, a}=\tfrac{\gamma}{\pi} \, \ln \tfrac{2 \, N}{{\cal M} \, L},
\end{equation}
where ${\cal M}$ denotes the physical mass of fermions (solitons) of the MT (SG) model, $a$ denotes the lattice constant, $L$ stands for the 
finite volume and $N$ is the number of lattice sites of the 6-vertex model, which should be even in this case. 
The relation of the anisotropy parameter $\gamma$ to the coupling constants of the quantum field theories 
(\ref{mTh_Lagrangian}) and (\ref{sG_Lagrangian})
is given by:
\begin{equation}\label{csatrel}
 \frac{\beta^2}{4 \pi}=\frac{1}{1+\tfrac{g}{4 \pi}}=2 (1-\tfrac{\gamma}{\pi}).
 \end{equation}
 For later convenience we also introduce a new parameterization for the anisotropy parameter:
\begin{equation} \label{csatrel1}
\gamma=\tfrac{\pi}{p+1}, \quad \text{with} \quad 0<p< \infty, \quad  \text{then:} \quad \frac{\beta^2}{4 \pi}=\frac{2 p}{p+1}.
\end{equation}
In the language of the new parameter $p,$ the $p=1$ point corresponds to the free-fermion point of the theory and 
the $0<p<1$ and $1<p$ regimes correspond to the attractive and repulsive regimes of the model, respectively.

\section{NLIE for the counting-function} \label{NLIEsect}

It is well-known, that the counting function satisfies a set of nonlinear integral 
equations (NLIE) \cite{KP1}-\cite{FRT3}. The main advantage of this set of equations is that 
 it has a well-defined continuum limit in the light-cone regularized picture \cite{ddv92}. 
Thus it is a suitable tool, through which the continuum limit of the counting-function 
can be defined. In this paper we present the equations only for the pure hole sector 
of the model, which was derived first in \cite{fioravanti}. The equations are of the form: 
\begin{equation} \label{DDVlat}
\begin{split}
&Z(\lambda)=N \pi \left( \chi_F(\lambda-\rho_0)+\chi_F(\lambda+\rho_0) \right)+2 \pi \sum\limits_{k=1}^{m_H} \, \chi(\lambda-h_k)+ \\
+ \, &\int\limits_{-\infty}^{\infty} \frac{d\lambda'}{ i} \, G(\lambda-\lambda'-i\eta) \,
 {\cal L}_+(\lambda'+i \eta)
-\, \int\limits_{-\infty}^{\infty} \frac{d\lambda'}{ i} \, G(\lambda-\lambda'+i\eta) \, {\cal L}_- (\lambda'-i \eta),
\end{split}
\end{equation}
where $0<\eta$ is a small contour deformation parameter, ${\cal L}_\pm(\lambda)$ stand for the nonlinear combinations:
\begin{equation} \label{calLpm}
\begin{split}
{\cal L}_\pm(\lambda)=\ln\left(1+(-1)^\delta \! e^{\pm i \, Z(\lambda )} \! \right),
\end{split}
\end{equation}
and the functions $\chi_F,\chi$ and $G$ are given by the formulas:
\begin{eqnarray}
\chi_F(\lambda)&=&\tfrac{1}{\pi} \arctan \left[ \tanh(\tfrac{\pi \lambda}{2 \gamma})\right], \label{chiF} \\
G(\lambda)&=&\int\limits_{-\infty}^{\infty} \frac{d\omega}{2 \pi} e^{-i \omega  x} \, 
\tilde{G}(\omega), \quad \qquad \quad \tilde{G}(\omega)=\frac{1}{2} 
\frac{\sinh(\tfrac{\pi \omega}{2}(1-\tfrac{2 \gamma}{\pi}))}
{\cosh(\tfrac{\gamma \omega}{2}) \sinh(\tfrac{\pi \omega}{2}(1-\tfrac{ \gamma}{\pi}))}, \label{Gdef} \\
\chi(\lambda)&=&\int\limits_0^\lambda d\lambda' \, G(\lambda'). \label{chidef}
\end{eqnarray}
In order for the equation (\ref{DDVlat}) to be complete, quantization equations for the holes should also be 
imposed:
\begin{equation} \label{BAEh}
Z({h}_k)=2 \pi \, I_k, \quad  \quad I_k\in \mathbb{Z}+\tfrac{1+\delta}{2}, \qquad k=1,..,m_H.
\end{equation}
The counting-function has a well-defined continuum limit \cite{ddv92,ddv95,fioravanti,ddv97,FRT1,FRT2,FRT3}, 
which is a simple 
$N \to \infty$ limit of the lattice counting-function (\ref{countfv}), 
such that the hole quantum numbers are kept fixed and $\rho_0$ is also tuned according to (\ref{rho}). 
We will use the notations $\hat{Z}(\lambda)$ and $\hat{\cal L}_\pm(\lambda)$ for the continuum limits of the functions 
${Z}(\lambda)$ and ${\cal L}_\pm(\lambda),$ respectively:
\begin{equation} \label{contfuns}
\begin{split}
\hat{Z}(\lambda)=\lim_{N \to \infty} Z(\lambda), \qquad \hat{\cal L}_\pm(\lambda)=\lim_{N \to \infty} 
{\cal L}_\pm(\lambda).
\end{split}
\end{equation}
The continuum limit of the counting-function satisfies the nonlinear integral equation as follows 
\cite{ddv92}\cite{Fevphd}:
\begin{equation} \label{DDVcont}
\begin{split}
&\hat{Z}(\lambda)=\ell \sinh(\tfrac{\pi}{\gamma}\lambda)+2 \pi \sum\limits_{k=1}^{m_H} \, \chi(\lambda-h_k)+ \\
+ \, &\int\limits_{-\infty}^{\infty} \frac{d\lambda'}{ i} \, G(\lambda-\lambda'-i\eta) \,
 \hat{\cal L}_+(\lambda'+i \eta)
-\, \int\limits_{-\infty}^{\infty} \frac{d\lambda'}{ i} \, G(\lambda-\lambda'+i\eta) \, \hat{\cal L}_- (\lambda'-i \eta),
\end{split}
\end{equation}
where $\ell= {\cal M}\, L$ with ${\cal M}$  being the physical mass of solitons (fermions) and with $L$ being the 
finite volume.  The holes are subjected to the continuum limit of the lattice quantization equations (\ref{BAEh}):
%continuum analogs of the 
%same type of quantization equations as their lattice counterparts: 
\begin{equation} \label{BAEhc}
\hat{Z}({h}_k)=2 \pi \, I_k, \quad  \quad I_k\in \mathbb{Z}+\tfrac{1+\delta}{2}, \qquad k=1,..,m_H.
\end{equation}

The importance of the counting function is that the positions of all Bethe-roots are encoded into it. 
In the Algebraic Bethe Ansatz framework all physical quantities can be expressed in terms of the Bethe-roots. 
Thus,  
it is natural to expect that all physical quantities can be expressed in terms of the counting-function, as well. 
The description of physical quantities in the language of the counting-function proves to be very useful, when 
the continuum limit should be taken.

\section{Gaudin-matrix for soliton states} \label{Gaudinsect}

The norm square of a Bethe-eigenstate (\ref{sajatvec}) is proportional to the determinant of the lattice 
Gaudin-matrix defined by (\ref{Phiab}). 
In this definition the derivatives of the counting-function with respect to the 
positions of the Bethe-roots should be taken. To be more precise $Z(\lambda)$ defined by (\ref{countfv}) should be considered as a function depending on the parameters $\lambda_a.$ In more detail the precise definition of 
(\ref{Phiab}) can be written as follows: 
\begin{equation} \label{Phiab1}
\begin{split}
\Phi_{ab}(\vec{\lambda})=-i \, \tfrac{\partial}{\partial x_a} \, Z(x_b|\vec{x}){\big|}_{\vec{x}=\vec{\lambda}},
\end{split}
\end{equation}
where $\vec{\lambda}=(\lambda_1,..,\lambda_m)$ with $\lambda_j$s being the solutions of (\ref{BAE}) and the function $Z(\lambda|\vec{x})$ is defined by a formula, which can be obtained from (\ref{countfv}) after the 
$\lambda_a \to x_a$ replacements:
\begin{equation} \label{countfvx}
Z(\lambda|\vec{x})=\frac{N}{2} \, \left( \phi_1(\lambda-\rho_0)+\phi_1(\lambda+\rho_0)\right)-\sum\limits_{k=1}^{m} \, \phi_2(\lambda-x_k).
\end{equation}
From the definitions (\ref{countfv}) and (\ref{countfvx}) it follows, that:
\begin{equation} \label{ZZL}
\begin{split}
Z(\lambda)=Z(\lambda|\vec{\lambda}).
\end{split}
\end{equation}
Formula (\ref{ZZL}) implies that $Z(\lambda)$ is a function of $m$ parameters, which are actually
the Bethe-roots of the eigenstate under consideration. In this context one think of the counting-function as 
a function characterized by $m+2$ parameters. These parameters are the number of lattice sites $N,$
 the inhomogeneity parameter $\rho_0$ and $\vec{\lambda};$ the vector containing the spectral parameters 
of the elementary excitations above the reference state $|0 \rangle.$  

The NLIE  for the counting function (\ref{DDVlat}) suggests another "parameterization" for the counting-function. 
In this description $Z(\lambda)$ seems to be the function of only $m_H+2$ parameters; the number of lattice sites $N,$
 the inhomogeneity parameter $\rho_0$ and the positions of the holes, which are the spectral parameters of the
elementary excitations above the true antiferromagnetic vacuum of the model. Thus one can define another Gaudin-matrix, 
which is defined by differentiating $Z(\lambda)$ with respect to the hole's positions:   
\begin{equation} \label{Qdef}
\begin{split}
Q_{jk}=\tfrac{\partial}{\partial h_j} \, Z(h_k), \qquad j,k=1,...,m_H.
\end{split}
\end{equation}
In the sequel we will call this matrix the dressed Gaudin-matrix, since its definition arises from the dressed 
excitations of the model. Taking the derivative of (\ref{DDVlat}) with respect to $h_j$ one obtains 
the following formula for $Q_{jk}:$
\begin{equation} \label{Qjk1}
\begin{split}
Q_{jk}=Z'(h_k) \, \delta_{jk}+{\cal G}_j(h_k), \qquad j,k=1,...,m_H,
\end{split}
\end{equation}
where the functions ${\cal G}_j(\lambda)$ are defined by the solutions of the linear integral equations 
as follows:
\begin{equation} \label{calGj}
\begin{split}
{\cal G}_j(\lambda)\!-\!\!\sum\limits_{\alpha=\pm}  \!\! \int\limits_{-\infty}^{\infty} \!\! d\!\lambda' \,
G(\lambda\!-\! \lambda'\!-\!i \,  \alpha \, \eta) \, {\cal F}_\alpha(\lambda'\!+\!i \, \alpha \, \eta)  
{\cal G}_j(\lambda'\!+\!i \, \alpha \, \eta)\!\!=\!\!-2 \pi \, G(\lambda\!-\!h_j), \quad j=1,..,m_H,
\end{split}
\end{equation}
where $h_j$ is the $j$th hole's position and the functions ${\cal F}_\pm(\lambda)$ are defined by:
\begin{equation} \label{calF}
{\cal F}_{\pm}(\lambda)=\frac{(-1)^\delta \, e^{\pm i \, Z(\lambda)}}
{1+(-1)^\delta \, e^{\pm i \, Z(\lambda)}}. 
\end{equation}
Formula (\ref{calGj}) implies, that the functions ${\cal G}_j(\lambda)$ originate from a single function of two arguments:
\begin{equation} \label{GjGt}
\begin{split}
{\cal G}_j(\lambda)=\tilde{\cal G}(\lambda,h_j), \quad j=1,..,m_H,
\end{split}
\end{equation}
such that $\tilde{\cal G}(\lambda,\lambda'')$ is defined by the solution of the linear integral equation 
as follows:
\begin{equation} \label{calGt}
\begin{split}
\tilde{\cal G}(\lambda,\lambda'')\!-\!\!\sum\limits_{\alpha=\pm}  \int\limits_{-\infty}^{\infty} \! d\!\lambda' \,
G(\lambda-\! \lambda'-i  \alpha  \eta) \, {\cal F}_\alpha(\lambda'+i \, \alpha \, \eta)  
\tilde{\cal G}(\lambda'+i \, \alpha \, \eta,\lambda'')\!\!=\!\!-2 \pi \, G(\lambda-\lambda''). 
\end{split}
\end{equation}
Thus the dressed Gaudin-matrix takes the form:
\begin{equation} \label{Qjk2}
\begin{split}
Q_{jk}=Z'(h_k) \, \delta_{jk}+\tilde{\cal G}(h_k,h_j), \qquad j,k=1,...,m_H.
\end{split}
\end{equation}
%%%%%%%%%%%%%%%%%%%%%%%%%%%%%%%%%%%%%%%%%%%%%%%%%%%%%%%%%%%%%%%%%%%%%%%%%%%%%%%%%%%%%
%%%%%%%%%%%%%%%%%%%%%%%%%%%%%%%%%%%%%%%%%%%%%%%%%%%%%%%%%%%%%%%%%%%%%%%%%%%%%%%%%%%%%
%%%%%%%%%%%%%%%%%%%%%%%%%%%%%%%%%%%%%%%%%%%%%%%%%%%%%%%%%%%%%%%%%%%%%%%%%%%%%%%%%%%%%
%%%%%%%%%%%%%%%%%%%%%%%%%%%%%%%%%%%%%%%%%%%%% Jo resz %%%%%%%%%%%%%%%%%%%%%%%%%%%%%%%%%%%%%%%%%%%%%
The continuum limit{\footnote{By continuum limit we always mean the $N \to \infty$ limit, such that $\rho_0$ is also
tuned according to (\ref{rho}).}} of $\tilde{G}(\lambda,\lambda'')$ also exists:
$\lim\limits_{N \to \infty} \tilde{\cal G}(\lambda,\lambda')=\tilde{\cal G}_c(\lambda,\lambda')$. 
From (\ref{calGt}) it follows, that it satisfies the linear integral equation as follows: 
\begin{equation} \label{calGtc}
\begin{split}
\tilde{\cal G}_c(\lambda,\lambda'')\!-\!\!\sum\limits_{\alpha=\pm}  \int\limits_{-\infty}^{\infty} \! d\!\lambda' \,
G(\lambda-\! \lambda'-i  \alpha  \eta) \, {\hat{\cal F}}_\alpha(\lambda'+i \, \alpha \, \eta)  
\tilde{\cal G}_c(\lambda'+i \, \alpha \, \eta,\lambda'')\!\!=\!\!-2 \pi \, G(\lambda-\lambda''), 
\end{split}
\end{equation}
where analogously to (\ref{contfuns}), $\hat{\cal F}_\pm(\lambda)$ stands for 
%we introduced the notation 
the continuum limit of ${\cal F}_\pm(\lambda):$
\begin{equation} \label{calFh}
\lim\limits_{N \to \infty}{\cal F}_\pm(\lambda)=\hat{\cal F}_\pm(\lambda).
\end{equation}
Then, in the continuum limit the dressed Gaudin-matrix takes the form:
\begin{equation} \label{Qjkc}
\begin{split}
{\hat{Q}}_{jk}=\hat{Z}'(h_k) \, \delta_{jk}+\tilde{\cal G}_c(h_k,h_j), \qquad j,k=1,...,m_H.
\end{split}
\end{equation}
%%%%%%%%%%%%%%%%%%%%%%%%%%%%%%%%%%%%%%%%%%%%%%%

For later convenience it is worth to represent the linear integral equations
(\ref{calGt})  and (\ref{calGtc}) in a form, where integrations run along the real axis. 
This can be done by "pushing" the integration contours on the real axis by taking the 
$\eta \to 0^+$ limit in these formulas. 
%(\ref{calGt})  and (\ref{calGtc}).
Exploiting that the functions $G(\lambda)$ and $\tilde{\cal G}(\lambda,\lambda')$ are regular around the 
real axis, the equations (\ref{calGt})  and (\ref{calGtc}) can be written in the alternative forms as follows:
\begin{equation} \label{calGt1}
\begin{split}
\tilde{\cal G}(\lambda,\lambda'')\!-\!\!  \int\limits_{-\infty}^{\infty} \! d\!\lambda' \,
G(\lambda-\! \lambda') \, \Omega_{\cal F}(\lambda')  
\tilde{\cal G}(\lambda',\lambda'')\!\!=\!\!-2 \pi \, G(\lambda-\lambda''), 
\end{split}
\end{equation}
\begin{equation} \label{calGt1c}
\begin{split}
\tilde{\cal G}_c(\lambda,\lambda'')\!-\!\!  \int\limits_{-\infty}^{\infty} \! d\!\lambda' \,
G(\lambda-\! \lambda') \, \Omega_{\hat{\cal F}}(\lambda') \, 
\tilde{\cal G}_c(\lambda',\lambda'')\!\!=\!\!-2 \pi \, G(\lambda-\lambda''), 
\end{split}
\end{equation}
where the functions $\Omega_{\cal F}(\lambda)$ and $\Omega_{\hat{\cal F}}(\lambda)$ are defined by the formulas as follows:
\begin{equation} \label{OMF}
\begin{split}
\Omega_{\cal F}(\lambda)=\lim_{\eta \to 0^+} \left( {\cal F}_+(\lambda+i \, \eta)+{\cal F}_-(\lambda-i \, \eta)
 \right),
\end{split}
\end{equation}
\begin{equation} \label{OMFc}
\begin{split}
\Omega_{\hat{\cal F}}(\lambda)=\lim_{\eta \to 0^+} \left( \hat{{\cal F}}_+(\lambda+i \, \eta)+\hat{{\cal F}}_-(\lambda-i \, \eta)
 \right).
\end{split}
\end{equation}
The representations (\ref{calGt1}) and (\ref{calGt1c}) will become important in section \ref{DETsect}, 
when the Gaudin-determinant is computed.

\section{Summation formulas} \label{SUMsect}

In the computations of the forthcoming sections we will need to compute sums of type as follows:
\begin{equation} \label{SUMf0}
\begin{split}
\Sigma_f=\sum\limits_{j=1}^m f(\lambda_j),
\end{split}
\end{equation}
where $f(\lambda)$ is an arbitrary function being regular in some neighborhood of the real axis and 
$\lambda_j$s are the roots of the Bethe-equations (\ref{BAE}).

In this section we derive integral expressions, which make it possible to eliminate 
the positions of the Bethe-roots from $\Sigma_f.$ Such formulas are very well-known 
in the literature \cite{ddv92}-\cite{Fevphd},\cite{En},\cite{En1}. The reason why we consider here 
these summation formulas, is to derive a special form for them, which proves to be very useful in the 
computations of the forthcoming sections.
%section \ref{DETsect}.

First we introduce a notation. In the sequel we will 
denote the positions of Bethe-roots and holes in common with $\hat{\lambda}_j:$
\begin{equation} \label{lambdahat}
\begin{split}
\{\hat{\lambda}_j\}_{j=1,..,m_\infty}=\{\lambda_j\}_{j=1,..,m} \cup \{h_j\}_{j=1,..,m_H}, \qquad m_\infty=m+m_H.
\end{split}
\end{equation}

\begin{figure}[htb]
\begin{flushleft}
%\vskip 10mm
\hskip 15mm
\leavevmode
\epsfxsize=120mm
\epsfbox{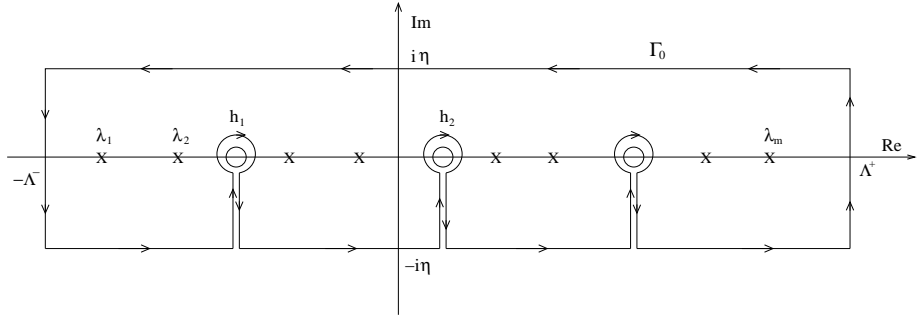}
%\vskip 10mm
\end{flushleft}
\caption{{\footnotesize
The figure represents the closed curve $\Gamma_0$ in (\ref{Sfderiv}). Crosses and circles represent the Bethe-roots and holes respectively. 
%The distance from the real axis is $\pm i \eta$ as indicated at the right hand side of the curve.
}}
\label{figG0}
\end{figure}

The starting point of the derivation is the recognition, that ${\cal F}_{\pm}(\lambda)$ defined by (\ref{calF}) 
have simple poles at the positions $\{\hat{\lambda}_j\}_{j=1,..,m_\infty}$ and otherwise they are regular in a 
small neighborhood of the real axis. From (\ref{BAEZ}) and (\ref{BAEH}) it can be shown, that for $\lambda \sim \hat{\lambda}_j$  they behave like: 
\begin{equation} \label{calFbehav}
\begin{split}
{\cal F}_\pm(\lambda)=\frac{\pm 1}{i \, Z'(\hat{\lambda}_j)} \frac{1}{\lambda-\hat{\lambda}_j}+O(1), \qquad 
j=1,..,m_\infty.
\end{split}
\end{equation}
It follows, that the residues at these points take the values:
\begin{equation} \label{rescalF}
\begin{split}
\text{Res}_{\lambda=\hat{\lambda}_j}{\cal F}_{\pm}(\lambda)=\frac{\pm 1}{i \, Z'(\hat{\lambda}_j)}, \qquad j=1,..,m_\infty.
\end{split}
\end{equation}
Then  the sum $\Sigma_f$ can be transformed into an integral expression in the following way:
\begin{equation} \label{Sfderiv}
\begin{split}
&\Sigma_f=\sum\limits_{j=1}^m \, f(\lambda_j)=\sum\limits_{j=1}^m f(\lambda_j) \, i \, Z'(\lambda_j) \, 
\text{Res}_{\lambda=\lambda_j}{\cal F}_{+}(\lambda)=\oint\limits_{\Gamma_0} \frac{d \lambda}{2 \pi} \, 
f(\lambda) \, Z'(\lambda) \, {\cal F}_+(\lambda)=\\
&=\int\limits_{-\Lambda^-}^{\Lambda^+} \frac{d \lambda}{2 \pi} f(\lambda-i \,\eta) 
Z'(\lambda-i \, \eta) {\cal F}_+(\lambda-i \, \eta)-\!\!\!\!
\int\limits_{-\Lambda^-}^{\Lambda^+} \frac{d \lambda}{2 \pi} f(\lambda+i \,\eta) 
Z'(\lambda+i \, \eta) {\cal F}_+(\lambda+i \, \eta) \\
&-\sum\limits_{j=1}^{m_H} f(h_j)+R_\Lambda=
-\sum\limits_{j=1}^{m_H} f(h_j)+R_\Lambda+\int\limits_{-\Lambda^-}^{\Lambda^+} \frac{d \lambda}{2 \pi} f(\lambda-i \,\eta) 
Z'(\lambda-i \, \eta) \\
&-\sum\limits_{\alpha=\pm} \int\limits_{-\Lambda^-}^{\Lambda^+} \frac{d \lambda}{2 \pi} \,
f(\lambda+i \alpha \,\eta) \,
Z'(\lambda+i \, \alpha \, \eta) \,  {\cal F}_\alpha(\lambda+i \, \alpha \, \eta),
\end{split}
\end{equation}
where the closed curve $\Gamma_0$ is depicted on figure \ref{figG0},
%%%%%%%%%%%%%%%%%%%%%%%%
%%%%%%%%%%%%%%%%%%%%%%%%% 
 $\Lambda^\pm$ are positive parameters satisfying the inequalities as follows:
\begin{equation} \label{Lpm}
\begin{split}
\text{max}_{ \!\!\!\! \!\!\!\!{\atop j}}\,\, \,\{\hat{\lambda_j}\}<\Lambda^+<\infty, \qquad
-\infty<-\Lambda^-<{\text{min}_{ \!\!\!\! \!\!\!\!{\atop j}}}\,\, \, \{\hat{\lambda}_j\}, 
\end{split}
\end{equation}
and $R_\Lambda$ denotes the "vertical" part of the integrations on $\Gamma_0:$
\begin{equation} \label{RL}
\begin{split}
{R}_\Lambda=i \sum\limits_{\alpha=\pm} \alpha \int\limits_{-\eta}^\eta \frac{d \tau}{2 \pi} 
f(\alpha \, \Lambda^\alpha+i \, \tau) \,
Z'(\alpha \, \Lambda^\alpha+i \, \tau)\, {\cal F}_+(\alpha \, \Lambda^\alpha+i \, \tau).
\end{split}
\end{equation}
In (\ref{Sfderiv}) we used (\ref{rescalF}) and the identity:
\begin{equation} \label{calFid}
\begin{split}
{\cal F}_+(\lambda-i \, \eta)=1-{\cal F}_-(\lambda-i \, \eta),
\end{split}
\end{equation}
which is a simple consequence of the definitions (\ref{calF}).
The maximal value of the positive contour deformation parameter $\eta$ is restricted by the 
singularity structure of the functions $f(\lambda), \, Z'(\lambda),$ and ${\cal F}_\pm(\lambda).$ 
Namely, it must be smaller, than the absolute value of the imaginary part of that singularity 
of these three functions, which lies the closest to the real axis. 
Shifting the contour to the real axis in the term $\int\limits_{-\Lambda^-}^{\Lambda^+} \frac{d \lambda}{2 \pi} f(\lambda-i \,\eta) Z'(\lambda-i \,\eta)$
 of (\ref{Sfderiv}) one obtains the final result 
as follows:
\begin{equation} \label{Sf1}
\begin{split}
\Sigma_f &= 
-\sum\limits_{j=1}^{m_H} f(h_j) 
+\!\!\! \int\limits_{-\Lambda^-}^{\Lambda^+} \frac{d \lambda}{2 \pi} f(\lambda) Z'(\lambda)
-\!\! \sum\limits_{\alpha=\pm} \int\limits_{-\Lambda^-}^{\Lambda^+} \frac{d \lambda}{2 \pi} \,
f(\lambda+i \alpha \,\eta) \,
Z'(\lambda+i \, \alpha \, \eta) \,  {\cal F}_\alpha(\lambda+i \, \alpha \, \eta)+ \\
&+ R_\Lambda+{R'}_\Lambda,
\end{split}
\end{equation}
where
\begin{equation} \label{RvL}
\begin{split}
{R'}_\Lambda=i \int\limits_{-\eta}^0 \frac{d \tau}{2 \pi} f(\Lambda^++i \, \tau) \,
Z'(\Lambda^++i \, \tau)-
i \int\limits_{-\eta}^0 \frac{d \tau}{2 \pi} f(-\Lambda^-+i \, \tau) \, Z'(-\Lambda^-+i \, \tau).
\end{split}
\end{equation}
For later convenience we will rephrase the integral representation (\ref{Sf1}) in two special forms.
%The choice between the two representations representations will be determined by the 
%large $\lambda$ asymptotics of the function $f(\lambda).$
The large $\lambda$ asymptotics of the function $f(\lambda)$ will determinate which from these two 
special representations should be used during the actual computations.

\underline{ \em Case I.}
This is the case, when the integrands of all integrals in (\ref{Sf1}) tend to zero at infinity in an integrable way.
Let us see, what kind of condition is imposed on the asymptotics of $f(\lambda)$ by this requirement.
From (\ref{countfv}) and (\ref{calF}) it can be seen that $Z'(\lambda)$ and ${\cal F}_\pm(\lambda)$ 
have the large $\lambda$ asymptotics as follows:
\begin{equation} \label{ZvcFasz}
\begin{split}
Z'(\lambda) \sim e^{-2 \, |\lambda|}, \qquad {\cal F_\pm} \sim \text{const}. 
\end{split}
\end{equation}
Then the condition, that all integrands in (\ref{Sf1}) should tend to zero at infinity in an integrable way, imposes 
the following requirement for the large $\lambda$ asymptotics of $f(\lambda):$ 
\begin{equation} \label{faszcond1}
\begin{split}
|f(\lambda)|\lessapprox \frac{e^{2 |\lambda|}}{|\lambda|^{1+\epsilon}}, \qquad 0<\epsilon \in {\mathbb R}.
\end{split}
\end{equation}
In this case the $\Lambda^\pm \to \infty$ limit can be taken in (\ref{Sf1}) and  
the vertical integrals (\ref{RL}) and (\ref{RvL}) tend to zero: $R_\Lambda \to 0, \, {R'}_\Lambda \to 0.$
Thus the formula (\ref{Sf1}) simplifies radically for functions with asymptotic behavior (\ref{faszcond1}):
\begin{equation} \label{Sfinf}
\begin{split}
\Sigma_f &= 
-\sum\limits_{j=1}^{m_H} f(h_j) 
+\!\!\! \int\limits_{-\infty}^{\infty} \frac{d \lambda}{2 \pi} f(\lambda) Z'(\lambda)
-\!\! \sum\limits_{\alpha=\pm} \int\limits_{-\infty}^{\infty} \frac{d \lambda}{2 \pi} \,
f(\lambda+i \alpha \,\eta) \,
Z'(\lambda+i \, \alpha \, \eta) \,  {\cal F}_\alpha(\lambda+i \, \alpha \, \eta).
\end{split}
\end{equation}
This formula will be used in the derivation of the formulas of section \ref{V0sect}.

\underline{Case II.} 
This is the case, when the large $\lambda$ asymptotics of $f(\lambda)$ does not satisfy the condition (\ref{faszcond1}). 
In this case the $\Lambda^\pm \to \infty$ limit cannot be taken in the formula (\ref{Sf1}). 
In section \ref{DETsect} we will need this case, too. For practical purposes, we rephrase (\ref{Sf1}) 
in a special form, in which the 
"vertical" terms $R_\Lambda,\, {R'}_\Lambda$ become zero. This is achieved by taking the $\eta \to 0^+$ limit 
in (\ref{Sf1}). The functions $f(\lambda)$ and $Z'(\lambda)$ are analytic around the real axis, this is why in the 
$\eta \to 0^+$ limit, the $f(\lambda \pm i \, \eta) \to f(\lambda)$ and $Z'(\lambda \pm i \, \eta) \to Z'(\lambda)$ 
replacements can be done in (\ref{Sf1}). On the other hand the situation is not so simple for 
${\cal F}_\pm(\lambda\pm i \, \eta ),$ because of their poles on the real axis (\ref{calFbehav}).
From (\ref{calF}) and (\ref{calFbehav}) the following small $\eta$ behavior can be derived:
\begin{equation} \label{calFeta}
\begin{split}
\sum\limits_{\alpha=\pm} {\cal F}_\alpha(\lambda+i \, \alpha \, \eta )=\Omega_{\cal F}^{(\eta)}(\lambda)+O(\eta), 
\end{split}
\end{equation}
where
\begin{equation} \label{OMeta}
\begin{split}
\Omega_{\cal F}^{(\eta)}(\lambda)=1-2 \pi \sum\limits_{j=1}^{m_\infty} \frac{\delta_\eta(\lambda-\hat{\lambda}_j)}{Z'(\hat{\lambda}_j)},
 \qquad \delta_\eta(\lambda)=\frac{\eta}{\pi} \frac{1}{\lambda^2+\eta^2}.
\end{split}
\end{equation}
The function $\delta_\eta(\lambda)$ is a smooth, analytic regularization of the Dirac-delta distribution $\delta(\lambda)$:
\begin{equation} \label{dd}
\begin{split}
\lim\limits_{\eta \to 0^+}\delta_\eta(\lambda)=\delta(\lambda).
\end{split}
\end{equation}
Formulas (\ref{calFeta})-(\ref{dd}) imply, that $\Omega_{\cal F}(\lambda)$ defined in (\ref{OMF}) can also be 
represented in the form as follows:
% admits the representation:
%Using (\ref{OMF}) and (\ref{calFeta}) $\Omega_{\cal F}(\lambda)$ admits the representation:
\begin{equation} \label{OM}
\begin{split}
\Omega_{\cal F}(\lambda)=1-2 \pi \sum\limits_{j=1}^{m_\infty} \frac{\delta(\lambda-\hat{\lambda}_j)}{Z'(\hat{\lambda}_j)}.
\end{split}
\end{equation}
We note that %$\Omega_{\cal F}(\lambda)$ and 
$\Omega_{\cal F}(\lambda)$ can be considered as a kernel of an
integral operator acting on the real axis. Due to its definition (\ref{OMF}) its action on 
a function $f$ with good-enough analytic properties can be written as an integration on 
appropriately shifted contours.  
Concretely:
\begin{equation} \label{OMFact}
\begin{split}
\int\limits_{-\infty}^\infty d \lambda \, f(\lambda) \, \Omega_{\cal F}(\lambda)=\sum\limits_{\alpha=\pm} \,
\int\limits_{-\infty}^\infty \!\! d \lambda \,
f(\lambda+i \,\alpha \, \eta) \, {\cal F}_\alpha(\lambda+i \, \alpha \, \eta),
\end{split}
\end{equation}
where on "good-enough" analytical properties we mean, that $f(\lambda)$
 is regular in a small neighborhood of the real axis and that the large $\lambda$ behavior of $f(\lambda)$ 
is such that the integrals converge on the right hand side of (\ref{OMFact}).

Now, we are in the position to bring (\ref{Sf1}) in case II to the form, which will be useful 
in the computations of section \ref{DETsect}:
\begin{equation} \label{Sfcut}
\begin{split}
\Sigma_f &= 
-\sum\limits_{j=1}^{m_H} f(h_j) 
+\!\!\! \int\limits_{-\Lambda^-}^{\Lambda^+} \frac{d \lambda}{2 \pi} f(\lambda) Z'(\lambda)
-\!\! \sum\limits_{\alpha=\pm} \int\limits_{-\Lambda^-}^{\Lambda^+} \frac{d \lambda}{2 \pi} \,
f(\lambda) \,
Z'(\lambda) \, \Omega_{\cal F}(\lambda).
\end{split}
\end{equation}
Since we would like to exploit the advantages of the $L^2$-function space, we rewrite (\ref{Sfcut}) 
into a form, which contains only smooth regular functions under the integrations,
\begin{equation} \label{Sfreg}
\begin{split}
\Sigma_f &= 
-\sum\limits_{j=1}^{m_H} f(h_j) 
+\!\!\! \int\limits_{-\Lambda^-}^{\Lambda^+} \frac{d \lambda}{2 \pi} f(\lambda) Z'(\lambda)
-\lim\limits_{\eta \to 0^+}
\! \sum\limits_{\alpha=\pm} \int\limits_{-\Lambda_\eta^-}^{\Lambda_\eta^+} \frac{d \lambda}{2 \pi} \,
f(\lambda) \,
Z'(\lambda) \, \Omega_{\cal F}^{(\eta)}(\lambda).
\end{split}
\end{equation}
Here we introduced the $\eta$ dependent cutoffs $\Lambda^{\pm}_\eta,$ which satisfy $\eta$ dependent 
 modifications of the inequalities (\ref{Lpm}):
\begin{equation} \label{Lpmreg}
\begin{split}
\text{max}_{ \!\!\!\! \!\!\!\!{\atop j}}\,\, \,\{\hat{\lambda_j}\}+\tfrac{1}{\sqrt{\eta}}<\Lambda_\eta^+<\infty, \qquad
-\infty<-\Lambda_\eta^-<{\text{min}_{ \!\!\!\! \!\!\!\!{\atop j}}}\,\, \, \{\hat{\lambda}_j\}-\tfrac{1}{\sqrt{\eta}}. 
\end{split}
\end{equation}
These $\eta$ dependent modifications of $\Lambda^\pm$ were introduced, so that the width of the 
regularized delta-functions 
corresponding to the maximal values of $\hat{\lambda}_j$ in (\ref{OMeta}) fall within the integration range and give 
contribution. Formulas (\ref{Sfcut}) and (\ref{Sfreg}) will play important role in the 
forthcoming sections, when the determinant of the Gaudin-matrix (\ref{Gm1}) is computed.

\section{The Gaudin-determinant} \label{DETsect}

In this section we rewrite the determinant of the Gaudin-matrix (\ref{Gm1}) into a form, which 
makes it possible to compute it in the continuum limit. As a first step we rewrite  ${\Phi}_{ab}$ 
of (\ref{Gm1}) in the form as follows:
\begin{equation} \label{PHIab}
\begin{split}
{\Phi}_{ab}=-i \, Z'(\lambda_b) \cdot \left( \delta_{ab} + U_{ab}\right), \qquad a,b=1,..m,
\end{split}
\end{equation} 
with
\begin{equation} \label{Uab}
\begin{split}
U_{ab}=U(\lambda_a,\lambda_b)=\frac{2 \pi \, K(\lambda_a-\lambda_b)}{Z'(\lambda_b)}, \qquad a,b=1,...,m,
\end{split}
\end{equation}
where $K(\lambda)$ is given by (\ref{K1}). Then $\text{det}_{ \!\!\!\! \!\!\!\!{\atop m}}\,\, \, \Phi$ 
can be written as a product of two terms:
\begin{equation} \label{detFi1}
\begin{split}
\text{det}_{ \!\!\!\! \!\!\!\!{\atop m}}\,\, \, \Phi=\Phi_0 \cdot \Phi_U,
%(-i)^m \, \prod\limits_{k=1}^m Z'(\lambda_k) \cdot 
%\text{det}_{ \!\!\!\! \!\!\!\!{\atop m}}\,\, \, (\delta_{ab}+U_{ab}).
\end{split}
\end{equation}
where 
\begin{equation} \label{trivial1}
\begin{split}
\Phi_0=(-i)^m \, \prod\limits_{k=1}^m Z'(\lambda_k),
\end{split}
\end{equation}
\begin{equation} \label{FiU}
\begin{split}
\Phi_U =\text{det}_{ \!\!\!\! \!\!\!\!{\atop m}}\,\, \, (\delta_{ab}+U_{ab}).
\end{split}
\end{equation}
The term $\Phi_0$ can be easily computed with the help of (\ref{Sfinf}). 
Nevertheless, we refrain from its explicit computation, since we will see in section \ref{SUMMARYsect}, that 
it cancels from the full expression of the norm $||\vec{\lambda}||.$

The other term $\Phi_U$ is a Fredholm-determinant. Its computation is the primary purpose of this section.
%\begin{equation} \label{FiU}
%\begin{split}
%\Phi_U =\text{det}_{ \!\!\!\! \!\!\!\!{\atop m}}\,\, \, (\delta_{ab}+U_{ab}).
%\end{split}
%\end{equation}
%Then
% \begin{equation} \label{detPhi2}
%\begin{split}
%\text{det}_{ \!\!\!\! \!\!\!\!{\atop m}}\,\, \,\Phi=\Phi_0 \cdot \Phi_U.
%\end{split}
%\end{equation}
For the computation of $\Phi_U$ we will use the Plemelj-formula \cite{Plem,Fred}, which expresses a Fredholm-determinant 
in terms of traces by the formula as follows:
\begin{equation} \label{Plemelj}
\begin{split}
\text{det} (1+A)=\exp\left\{\sum\limits_{n=1}^\infty \tfrac{(-1)^{n+1}}{n} \, \text{Tr} (A^n) \right\},
\end{split}
\end{equation}
where $\text{Tr}$ stands for trace. The formula (\ref{Plemelj}) is convergent if 
\begin{equation} \label{Plemconv}
|\lambda_1(A)|<1,
\end{equation}
 where $\lambda_1(A)$ stands for the largest eigenvalue of the operator $A.$ 
To check whether a Plemelj-series (\ref{Plemelj}) is convergent, $|\lambda_1(A)|$ should be computed.
In our cases the following formula proves to be useful for its determination:
 \begin{equation} \label{l1A}
\begin{split}
|\lambda_1(A)|=\lim\limits_{n \to \infty}\exp\left\{ \tfrac{1}{n} \ln |\text{Tr}(A^n)| \right\}.
\end{split}
\end{equation}
In the forthcoming subsections we compute $\Phi_U$ by the appropriate 
application of the Plemelj-formula (\ref{Plemelj}).
%Now, we apply the Plemelj-formula (\ref{Plemelj}) for our Fredholm-determinant (\ref{FiU}).

\subsection{The determination of $\Phi_U$}

To avoid extra difficulties, during the computations we will assume, that
 $Z'(\lambda_j)>0$ for all $j=1,...m.$ 
In order to apply the Plemelj-formula for $\Phi_U$ (\ref{FiU}), one has to 
know, whether it converges.  Unfortunately, we cannot prove, that the
Plemelj-formula for $\Phi_U$ would be convergent in the whole allowed range of the anisotropy 
parameter ($0<\gamma<\pi.$) Nevertheless, we will show in the forthcoming 
lines, that there is a region in the anisotropy parameter, where the series is convergent for sure. 
We will compute $\Phi_U$ in this convergent regime and the result outside of this regime 
will be obtained by analytical continuation in the anisotropy parameter. 

To compute $\Phi_U$ with the help of (\ref{Plemelj}) one needs to compute traces of the powers of $U_{ab}$.
Using (\ref{Uab}) and (\ref{Kprop1}) it is easy to find a majorant for $|\text{Tr}(U^n)|:$
\begin{equation} \label{trU1}
\begin{split}
|\text{Tr}(U^n)|\leq\sum\limits_{a_1,..,a_n=1}^m |U(\lambda_{a_1},\lambda_{a_2})....U(\lambda_{a_n},\lambda_{a_1})| \leq (2 \pi \, |K(0)| \, \Sigma_m)^n,
\end{split}
\end{equation}
where
\begin{equation} \label{Sm}
\begin{split}
\Sigma_m=\sum\limits_{j=1}^m \frac{1}{Z'(\lambda_j)},
\end{split}
\end{equation}
which is a positive number due to the assumption $Z'(\lambda_j)>0.$
From (\ref{trU1}) it follows, that the Plemelj-series for $\Phi_U$ surely converges if one works in the region of the anisotropy 
parameter, where the following inequality holds:
\begin{equation} \label{ineqU}
\begin{split}
2 \pi |K(0)| \, \Sigma_m<1.
\end{split}
\end{equation}
This condition is obviously satisfied if the value of the anisotropy is close enough to the free-fermion point. 
This statement is a simple consequence of the formula (\ref{Kzero}) and the fact that $\Sigma_m\neq 0$ at the $p=1$ free-fermion point. 
In the sequel we will work in this region of $p$, where the Plemelj-formula 
for $\Phi_U$ is convergent.

Our aim is to rephrase $\Phi_U$ in a form which is appropriate for taking the continuum limit. 
This is why we should somehow turn from discrete 
variables to continuous ones. In (\ref{trU1}) such sums arise for which we have derived the summation formulas (\ref{Sfinf}) and (\ref{Sfreg}). 
This is why our strategy is to rewrite $\text{Tr}(U^n)$ as the trace of the $n$th power of an appropriate integral operator.
To do so, first one has to specify the function space or Hilbert-space on which the integral operators act. Formula (\ref{Sfreg}) 
suggest the Hilbert-space ${\cal H}$ to be $L^2(-\Lambda^-_\eta,\Lambda^+_\eta).$ 
Then consider the integral operator $\hat{U}_\eta$ acting on ${\cal H},$ with the kernel:
\begin{equation} \label{Ueta}
\begin{split}
U_\eta(\lambda,\lambda')=K(\lambda-\lambda')-2 \pi \sum\limits_{j=1}^{m_H} \frac{K(\lambda-\lambda') \, \delta_\eta(\lambda'-h_j)}{Z'(h_j)}-
K(\lambda-\lambda') \, \Omega_{\cal F}^{(\eta)}(\lambda'),
\end{split}
\end{equation}
where the functions $K,$ $\delta_\eta$ and $\Omega_{\cal F}^{(\eta)}$ are defined  in (\ref{K1}) and (\ref{OMeta}). 
Using (\ref{OMeta}), the kernel $U_\eta(\lambda,\lambda')$ can be rephrased from (\ref{Ueta}) in an equivalent 
form being closer in nature to the original discrete problem:
\begin{equation} \label{U0eta}
\begin{split}
U_\eta(\lambda,\lambda')=
2 \pi \sum\limits_{j=1}^{m} \frac{K(\lambda-\lambda') \, \delta_\eta(\lambda'-\lambda_j)}{Z'(\lambda_j)}.
\end{split}
\end{equation}
From (\ref{U0eta}) the following trace relation can be shown: 
\begin{equation} \label{TRrel1}
\begin{split}
\text{Tr}(U^n)=\lim\limits_{\eta \to 0^+}\text{Tr}(\hat{U}_\eta^n), \qquad  1 \leq n \in {\mathbb Z},
\end{split}
\end{equation}
where 
\begin{equation} \label{TrUeta}
\begin{split}
\text{Tr}(\hat{U}_\eta^n)=\!\!\! \int\limits_{-\Lambda^-_\eta}^{\Lambda^+_\eta} \!\!\! d\lambda_1 ...d\lambda_n    \, U_\eta(\lambda_1,\lambda_2)\, U_\eta(\lambda_2,\lambda_3)...
U_\eta(\lambda_n,\lambda_1).
\end{split}
\end{equation}
From (\ref{U0eta}) and from the representation (\ref{TrUeta}), one can also show, that:
\begin{equation} \label{TrUetamaj}
\begin{split}
|\text{Tr}(\hat{U}_\eta^n)|\leq  (2 \pi \, |K(0)| \, \Sigma_m)^n.
\end{split}
\end{equation}
This formula together with (\ref{trU1}) implies, that in the region (\ref{ineqU}) 
in which we work the Plemelj-formula for $\text{det}(1+\hat{U}_\eta)$ will converge, too.
The convergence of the Plemelj-series and  
the trace relation (\ref{TRrel1}) implies, that:
\begin{equation} \label{TRrel2}
\begin{split}
\Phi_U =\text{det}_{ \!\!\!\! \!\!\!\!{\atop m}}\,\, \, (\delta_{ab}+U_{ab}) = \lim\limits_{\eta \to 0^+} \text{det} (1+\hat{U}_\eta).
\end{split}
\end{equation}
%%%%%%%%%%%%%%%%%%%%%%%%%%%%%%%%%%%%%%%%%%%%%%%%%%%%%%%%%%%%%%%%%%%%%%%%%%%%%%%%%%%%%%%%%%%%%%%%%%%%
One can recognize, that in the definition (\ref{Ueta}) of $\hat{U}_\eta$ an $\eta$-dependent regularization of the 
Dirac-delta distribution arises. Since at the end of the computations the $\eta \to 0^+$ limit must be taken, one 
might ask why we cannot work directly in the $\eta \to 0^+$ limit, where Dirac-delta functions would enter the 
definition of this integral operator. 
The reason is that the $\eta$-deformed operator $\hat{U}_\eta$ has advantageous properties on ${\cal H}.$ 
Namely, it is a bounded, trace-class operator on ${\cal H}$ and this property is exploited in the subsequent 
computations{\footnote{All definitions and theorems concerning Fredholm-determinants and  trace-class operators, which 
we use in our actual computations can be found in the introduction of 
\cite{FBorn} and in the books \cite{BarryBook,Lax}}}. The operator defined directly in the $\eta \to 0^+$ limit: 
$\hat{U}=\lim\limits_{\eta \to 0^+} \hat{U}_\eta$, would not be a bounded  trace-class operator.

%%%%%%%%%%%%%%%%%%%%%%%%%%%%%%%%%%%%%%%%%%%%%%%%%%%%%%%%%%%%%%%%%%%%%%%%%%%%%%%%%%%%%%%%%%%%%%%%%%%%
%The reason for the $\eta$-regularization in the definition of $\hat{U}_\eta$ in (\ref{Ueta}) is 
%that in this case the integral operator 
%$\hat{U}_\eta$ is a bounded trace-class operator on ${\cal H},$ and this trace-class property 
%will be exploited in the subsequent computations.
%We just note, that the operator defined directly in the $\eta \to 0^+$ limit: 
%$\hat{U}=\lim\limits_{\eta \to 0^+} \hat{U}_\eta$, would not be a trace-class operator. 

The next step of our computations is to perform the dressing procedure in $\Phi_U.$ 
In the original formula for $U_{ab}$ the function $K(\lambda)$ arises, which is the derivative of the scattering-phase of elementary excitations above the completely 
ferromagnetic "bare" vacuum $|0\rangle$. Now, we rewrite $\Phi_U$ into a form, where the 
derivative of the scattering-phase of the elementary excitations above the 
true antiferromagnetic vacuum will arise. The procedure is quite similar to what happens 
in \cite{ddv92,ddv95} at the derivation of the NLIE (\ref{DDVlat}). 

For the sake of simplicity, we rewrite $\hat{U}_\eta$ in (\ref{Ueta}) into a form, which emphasize more, that it is a  
sum of three different type of terms:
\begin{equation} \label{Ueta3}
\begin{split}
\hat{U}_\eta=\hat{K}+\hat{H}_\eta+\hat{S}_\eta,
\end{split}
\end{equation}
where the kernels of these integral operators are given by:
\begin{eqnarray}
\hat{K} &\to& K(\lambda-\lambda'), \label{Khat}\\
\hat{H}_\eta &\to& H_\eta(\lambda,\lambda')=-2 \pi \sum\limits_{j=1}^{m_H} \frac{K(\lambda-\lambda') \, \delta_\eta(\lambda'-h_j)}{Z'(h_j)}, \label{Heta}\\
\hat{S}_\eta &\to& S_\eta(\lambda,\lambda')=
-K(\lambda-\lambda') \, \Omega_{\cal F}^{(\eta)}(\lambda'). \label{Seta}
\end{eqnarray}
In this notation $H_\eta$ and $S_\eta$ correspond to the contributions of the holes and the sea of Bethe-roots respectively.

Now, we make some trivial transformations on $\text{det} (1+\hat{U}_\eta).$ 
The only thing we exploit is that, for trace-class operators the determinant of the convolution product of 
two operators is equal to the product of the individual determinants \cite{BarryS},
\begin{equation} \label{detat}
\begin{split}
\text{det}&(1+\hat{U}_\eta)=\text{det} (1+\! \hat{K}+\hat{H}_\eta+\!\hat{S}_\eta)
=\text{det}\!\left(\!(1\!+\!\hat{K}) \! \! \star \!\! \left[ 1+\!(1+\hat{K})^{-1} \!\! \star \! \hat{H}_\eta+\!(1\!+\!\hat{K})^{-1} \! \star \! \hat{S}_\eta  \right]  \right)=\\
=\text{det}&(1+\hat{K}) \, \text{det} (1-\hat{A}_\eta-\hat{B}_\eta)=\text{det}(1+\hat{K}) \, \text{det}(1-\hat{B}_\eta) \, 
\text{det} \left(1-(1-\hat{B}_\eta)^{-1}\! \star \! \hat{A}_\eta\right),
\end{split}
\end{equation}
where we introduced the operators:
\begin{equation} \label{ABhat}
\begin{split}
\hat{A}_\eta=-(1+\hat{K})^{-1} \!\! \star \! \hat{H}_\eta, \qquad \hat{B}_\eta=-(1+\hat{K})^{-1} \!\! \star \! \hat{S}_\eta.
\end{split}
\end{equation}
It is worth to introduce another operator $\hat{\cal G}_\eta$ by the definition:
\begin{equation} \label{cGeta}
\begin{split}
\hat{\cal G}_\eta=(1-\hat{B}_\eta)^{-1}\! \star \! \hat{A}_\eta.
\end{split}
\end{equation}
With these definitions the result of (\ref{detat}) can be written as follows:
\begin{equation} \label{detres1}
\begin{split}
\text{det}(1+\hat{U}_\eta)=\text{det}(1+\hat{K}) \, \text{det}(1-\hat{B}_\eta) \, \text{det}(1-\hat{\cal G}_\eta).
\end{split}
\end{equation}
After these quite formal computations one should determine the kernels corresponding to the operators $\hat{A}_\eta, \hat{B}_\eta$ and $\hat{\cal G}_\eta.$
From definitions (\ref{ABhat}) one obtains:
\begin{eqnarray}\label{Aeta}
A_\eta(\lambda,\lambda') &=& 2 \pi \sum\limits_{j=1}^{m_H} \frac{G_\Lambda(\lambda,\lambda') \, \delta_\eta(\lambda'-h_j)}{Z'(h_j)},
\\ \label{Beta}
B_\eta(\lambda,\lambda') &=& G_\Lambda(\lambda,\lambda') \, \Omega_{\cal F}^{(\eta)}(\lambda'),
\end{eqnarray}
where $G_\Lambda(\lambda,\lambda')$ is the kernel of the operator $\hat{G}_\Lambda=(1+\hat{K})^{-1} \!\star  \hat{K}.$ 
As a consequence, it satisfies the linear integral equation as follows:
\begin{equation} \label{GL}
\begin{split}
G_\Lambda(\lambda,\lambda')\!+\! \int\limits_{-\Lambda_{\eta}^-}^{\Lambda_{\eta}^+} dz \, K(\lambda-z) \, G_\Lambda(z,\lambda')=K(\lambda-\lambda'), \qquad 
\lambda, \lambda' \in (-\Lambda_{\eta}^-,\Lambda_{\eta}^+).
\end{split}
\end{equation}
This kernel simplifies in the $\Lambda_\eta^\pm \to  \infty$ limit{\footnote{Since in this limit (\ref{GL}) 
can be solved in Fourier-space and the solution becomes $G(\lambda)$ of (\ref{Gdef}).}}:
\begin{equation} \label{GLlim}
\begin{split}
\lim\limits_{\Lambda_\eta^\pm \to \pm \infty} G_\Lambda(\lambda,\lambda')=G(\lambda-\lambda'),
\end{split}
\end{equation}
with $G(\lambda)$ given in  (\ref{Gdef}) and it is just the derivative of the soliton-soliton scattering phase.
We note, that (\ref{detres1}) with formulas (\ref{Aeta}), (\ref{Beta}) and (\ref{GL}) account for the dressing procedure, since 
this is a formula in which one can switch from the scattering-phase of "bare"-particles to that of the real physical particles.
From the series representation $G_\Lambda=\sum\limits_{n=1}^\infty \, (-1)^{n+1} (\star K)^n$ 
obtained by solving (\ref{GL}), and from 
(\ref{FK}) and (\ref{KoK0}) it can be shown, 
that the kernel $G_\Lambda$ is a bounded function, which becomes zero at the free-fermion point:
\begin{equation} \label{GLbound}
\begin{split}
|G_\Lambda(\lambda,\lambda')|\leq \int\limits_{-\infty}^{\infty} \!\! \frac{d \omega}{2 \pi} \frac{|\tilde{K}(\omega)|}{1-|\tilde{K}(\omega)|}=g(p), 
\qquad g(1)=0, \qquad \lambda,\lambda'\in (-\Lambda_\eta^-,\Lambda^+_\eta).
\end{split}
\end{equation}
 From (\ref{cGeta}), (\ref{Aeta}) and (\ref{Beta}) it follows, that the kernel of the operator $\hat{\cal G}_\eta$ satisfies the linear integral equation as follows:
\begin{equation} \label{cGllie}
\begin{split}
{\cal G}_\eta(\lambda,\lambda')\!- \!\!\!\! \! \int\limits_{-\Lambda_\eta^-}^{\Lambda^+_\eta} \!\!\!\! dz \, G_\Lambda(\lambda,z) \, \Omega_{\cal F}^{(\eta)}(z)\, 
{\cal G}_\eta(z,\lambda')\!=\!2  \pi \, G_\Lambda(\lambda,\lambda') \sum\limits_{j=1}^{m_H} \frac{\delta_\eta(\lambda'\!-\!h_j)}{Z'(h_j)}, \quad \,
\lambda, \lambda' \in (-\Lambda_{\eta}^-,\Lambda_{\eta}^+).
\end{split}
\end{equation}
 It is worth to search the solution of equation (\ref{cGllie}) in the form as follows:
 \begin{equation} \label{bG}
\begin{split}
{\cal G}_\eta(\lambda,\lambda')=\bar{\cal G}_\eta(\lambda,\lambda') \, \sum\limits_{j=1}^{m_H} \, \frac{\delta_\eta(\lambda'-h_j)}{Z'(h_j)}.
\end{split}
\end{equation}
 Then $\bar{\cal G}_\eta(\lambda,\lambda')$ should satisfy the equation:
\begin{equation} \label{bGllie}
\begin{split}
\bar{\cal G}_\eta(\lambda,\lambda')-\!\! \! \int\limits_{-\Lambda_\eta^-}^{\Lambda^+_\eta} \!\!\! dz \, G_\Lambda(\lambda,z) \, \Omega_{\cal F}^{(\eta)}(z)\, 
\bar{\cal G}_\eta(z,\lambda')=2  \pi \, G_\Lambda(\lambda,\lambda'), \qquad \lambda, \lambda' \in (-\Lambda_{\eta}^-,\Lambda_{\eta}^+).
\end{split}
\end{equation}
It is also important to introduce the $\eta \to 0^+$ limit of $\bar{\cal G}_\eta(\lambda,\lambda').$ It satisfies the linear integral equation as follows:
\begin{equation} \label{bGl1}
\begin{split}
\bar{\cal G}(\lambda,\lambda')-\!\! \! \int\limits_{-\Lambda^-}^{\Lambda^+} \!\!\! dz \, G_{\Lambda_0}(\lambda,z) \, \Omega_{\cal F}(z)\, 
\bar{\cal G}(z,\lambda')=2  \pi \, G_{\Lambda_0}(\lambda,\lambda'), \qquad \bar{\cal G}(\lambda,\lambda')=\lim\limits_{\eta \to 0^+} \bar{\cal G}_\eta(\lambda,\lambda'),
\end{split}
\end{equation}
where $G_{\Lambda_0}(\lambda,\lambda')=\lim\limits_{\eta \to 0^+} G_{\Lambda}(\lambda,\lambda').$
The linear equation (\ref{GL}) implies, that $G_\Lambda(\lambda,\lambda')$ tends to zero exponentially for large values of $\lambda'.$ 
As a consequence the integration can be extended to the whole real axis in (\ref{bGl1}) and using (\ref{OMF}) 
and (\ref{OMFact}), equation (\ref{bGl1}) can be rephrased in the form as follows:
\begin{equation} \label{bGl2}
\begin{split}
\bar{\cal G}(\lambda,\lambda')-\!\! \! \sum\limits_{\alpha=\pm} \int\limits_{-\infty}^{\infty} \! dz \,\, G_{\Lambda_0}(\lambda,z+i \, \alpha \, \eta) \, {\cal F}_\alpha(z+i \, \alpha \, \eta)\, 
\bar{\cal G}(z+i \, \alpha \, \eta,\lambda')=2  \pi \, G_{\Lambda_0}(\lambda,\lambda').
\end{split}
\end{equation}
Now, we are in the position to compute the three determinants in the right hand side of (\ref{detres1}) in 
the $\eta \to 0^+$ limit{\footnote{ We just recall, that as a consequence of (\ref{TRrel2}) taking the 
$\eta \to 0^+$ limit is necessary to obtain the norm of the Bethe wave-functions.}}.

\subsubsection{Computing $\lim\limits_{\eta \to 0^+} \text{det} (1-\hat{\cal G}_\eta)$}

For small enough values of $p-1,$ the determinant $\lim\limits_{\eta \to 0^+} \text{det} (1-\hat{\cal G}_\eta)$ can be represented by its convergent 
Plemelj-series. This representation allows one to rewrite this functional determinant as a finite dimensional one after the simple transformations as follows:
\begin{equation} \label{detcG1}
\begin{split}
&\lim\limits_{\eta \to 0^+} \ln \text{det} (1-\hat{\cal G}_\eta)=\lim\limits_{\eta \to 0^+} \sum\limits_{n=1}^\infty \frac{-1}{n}\,  \text{Tr}(\hat{\cal G}_\eta^n)=
 \sum\limits_{n=1}^\infty \frac{-1}{n} \lim\limits_{\eta \to 0^+} \,  \text{Tr}(\hat{\cal G}_\eta^n)= \\
 &=\sum\limits_{n=1}^\infty \frac{-1}{n} \lim\limits_{\eta \to 0^+} \!\! \int\limits_{-\Lambda^-_\eta}^{\Lambda^+_\eta} \!\!\! d\lambda_1...d\lambda_n \,\,\, {\cal G}_\eta(\lambda_1,\lambda_2)...
 {\cal G}_\eta(\lambda_n,\lambda_1)= \\
 &=\sum\limits_{n=1}^\infty \frac{-1}{n} \lim\limits_{\eta \to 0^+} \!\! \int\limits_{-\Lambda^-_\eta}^{\Lambda^+_\eta} \!\!\! d\lambda_1...d\lambda_n \,\,\, 
 \bar{\cal G}_\eta(\lambda_1,\lambda_2)...
 \bar{\cal G}_\eta(\lambda_n,\lambda_1) \, \sum\limits_{j_1=1}^{m_H}...\sum\limits_{j_n=1}^{m_H} \prod\limits_{k=1}^{m_H} \frac{\delta_\eta(\lambda_k-h_{j_k})}{Z'(h_{j_k})}
 = \\
 &=\sum\limits_{n=1}^\infty \frac{-1}{n} \, \sum\limits_{j_1=1}^{m_H}...\sum\limits_{j_n=1}^{m_H} \frac{\bar{\cal G}(h_1,h_2)...\bar{\cal G}(h_n,h_1)}{Z'(h_1)...Z'(h_n)}=
 \ln \text{det}_{ \!\!\!\! \!\!\!\!{\atop {m_H}}} \!\!\! \left[\delta_{jk}-\frac{\bar{\cal G}(h_j,h_k)}{Z'(h_k)} \right].
\end{split}
\end{equation}
In (\ref{detcG1}) we used the Plemelj-formula (\ref{Plemelj}) and (\ref{bG}). Furthermore we exploited, that in the 
$\eta \to 0^+$ limit: $\bar{\cal G}_\eta \to \bar{\cal G},$ $\delta_\eta \to \delta.$ This made it possible to evaluate 
all the integrals in (\ref{detcG1}) and to obtain the Plemelj-series of the 
determinant of an $m_H \times m_H$ matrix.

\subsubsection{Computing $\text{det} (1+\hat{K})$}

In the preceding pages we rewrote the Gaudin-determinant into a form, in which it is expressed in terms of the 
counting-function and the positions of the holes. %over the antiferromagnetic ground state of the model. 
%instead of the original description given in the language of Bethe-roots.  
%To turn from the lattice Bethe-roots to the holes over the antiferromagnetic ground state, 
Such a description proves to be useful, if one is interested in the large $N$ or continuum limit of the lattice model. 
In this limit, the positions of the minimal and maximal Bethe-roots or holes, which determine the 
cutoffs $\Lambda^\pm$ through (\ref{Lpm}), become of order $\ln N$ (i.e. $\Lambda_\pm \sim \pm 2 \ln N$).
This is why, we compute $\text{det} (1+\hat{K})$ in the limit, when $\Lambda^\pm$ is large and we neglect 
corrections, which tend to zero in the $\Lambda^\pm \to \infty$ limit. 
This can be done by applying the Plemelj-formula (\ref{Plemelj}) to $\text{det} (1+\hat{K}).$
\begin{equation} \label{1pK1}
\begin{split}
\ln \text{det} (1+\hat{K})=\sum\limits_{n=1}^{\infty} \frac{(-1)^{n+1}}{n} \, \text{Tr}(\hat{K}^n).  
\end{split}
\end{equation}
The necessary traces can be computed by applying formula (\ref{trO}):
\begin{equation} \label{trKn1}
\begin{split}
\text{Tr}(K^n)=\! \!\int\limits_{-\Lambda^-}^{\Lambda^+} \!\!\!\! d\lambda \,\, (\star K)^n(\lambda,\lambda),
\end{split}
\end{equation}
where
\begin{equation} \label{KnL}
\begin{split}
(\star K)^n(\lambda,\lambda)=\!\! \int\limits_{\Lambda^-}^{\Lambda^+} \!\! d\lambda_2...d\lambda_n \,
K(\lambda-\lambda_2) \, K(\lambda_2-\lambda_3)...K(\lambda_n-\lambda). 
\end{split}
\end{equation}
It is worth to represent the integration range in (\ref{KnL}) as follows:
$\int\limits_{-\Lambda^-}^{\Lambda^+}=\int\limits_{-\infty}^{\infty}-\int\limits_{-\infty}^{-\Lambda^-}-
\int\limits_{\Lambda^+}^{\infty}.$ 
Then the leading in large $\Lambda^\pm$ contribution comes from the terms, when all integrations run from 
$-\infty$ to $\infty.$ This term can be easily computed using the Fourier-transform (\ref{FK}). 
The large $\lambda$ asymptotics: $K(\lambda)\sim e^{-2 |\lambda|},$ implies that 
the corrections in (\ref{KnL}) are of order $e^{-4|\Lambda^\pm|}.$ 
Thus the large $\Lambda^\pm$ result for $(\star K)^n(\lambda,\lambda)$ can be summarized by the formula:
 \begin{equation} \label{Knl1}
\begin{split}
(\star K)^n(\lambda,\lambda)=\!\! \int\limits_{-\infty}^{\infty} \!\! \frac{d\omega}{2 \pi} \, \tilde{K}(\omega)^n+
O(e^{-4|\Lambda^\pm|}).
\end{split}
\end{equation}
Then the $\lambda$ integration can be evaluated in (\ref{trKn1}) admitting the following result for the trace:
\begin{equation} \label{trKn2}
\begin{split}
\text{Tr}(K^n)=\! \!(\Lambda^++\Lambda^-) \,\int\limits_{-\infty}^{\infty} \!\! \frac{d\omega}{2 \pi} \, 
\tilde{K}(\omega)^n \left[1+O(e^{-4|\Lambda^\pm|}) \right].
\end{split}
\end{equation}
Inserting (\ref{trKn2}) into (\ref{1pK1}) the sum can be evaluated, and one ends up with the final result as follows:
\begin{equation} \label{det1pK}
\begin{split}
\text{det}(1+\hat{K})=\exp\left\{ 
\Lambda \,\int\limits_{-\infty}^{\infty} \!\! \frac{d\omega}{2 \pi} \, 
\ln(1+\tilde{K}(\omega))+ O(\Lambda \,e^{-4|\Lambda^\pm|})
\right\},
\end{split}
\end{equation} 
where for short we introduced the notation $\Lambda=\Lambda^+\!+\!\Lambda^-.$

\section{ $\Phi_U$ in the continuum limit} \label{FIUsect}

In this section we discuss the continuum limit of the formula (\ref{detres1}) in the $\eta \to 0^+$ limit. 
The formula under consideration is a product of three terms. The continuum limit of each term can be discussed   
separately. The simplest term is the determinant $\text{det}(1+\hat{K}).$ In the continuum limit 
the widest Bethe-roots or holes (Bethe-objects) tend to plus or minus infinity: $\Lambda^\pm \to \pm \infty$.
Thus, formula (\ref{det1pK}) implies, that $\text{det}(1+\hat{K})$ is either divergent or tend to zero 
in the $N \to \infty$ limit.
\begin{equation} \label{det1pKc}
\begin{split}
\text{det}(1+\hat{K})\stackrel{N \to \infty}{\to}\exp\left\{ 
\Lambda_c \,\int\limits_{-\infty}^{\infty} \!\! \frac{d\omega}{2 \pi} \, 
\ln(1+\tilde{K}(\omega))
\right\},
\end{split}
\end{equation} 
where the singular behavior of the continuum limit is governed by $\Lambda_c,$ which is 
given by the range of Bethe-roots and holes: 
\begin{equation} \label{Lcont}
\begin{split}
\Lambda_c=\text{max}_{ \!\!\!\! \!\!\!\!{\atop j}}\,\, \, \,\hat{\lambda}_j-\text{min}_{ \!\!\!\! \!\!\!\!{\atop j}}\,\, \,\, \hat{\lambda}_j.
\end{split}
\end{equation}
From the large $N$ solution of the quantization equations $Z(\hat{\lambda}_j)=2 \pi \, \hat{I}_j,$ 
it turns out that:
\begin{equation} \label{Lcont1}
\begin{split}
\Lambda_c=4 \, \ln N+O(1).
\end{split}
\end{equation}
This implies, that the large $N$ behavior of $\text{det}(1+\hat{K})$ is under control. It has 
a power behavior in $N,$ such that the power is determined by the integral term in the exponent of 
(\ref{det1pKc}). From the Fourier-transform (\ref{FK}) it follows, that this power is negative 
in the attractive regime ($0<p<1$) and positive in the repulsive regime ($1<p$). Thus in the attractive regime 
$\text{det}(1+\hat{K})$ tend to zero, while in the repulsive regime it diverges.

The determination of the continuum limit of $\lim\limits_{\eta \to 0^+} \text{det} (1-\hat{\cal G}_\eta)$ is 
a simple task, since in (\ref{detcG1}) it has been reduced to the determinant of an $m_H \times m_H$ matrix. 
Thus, the continuum limit procedure is simply to take the continuum limit of the matrix elements of this finite
matrix. Comparing (\ref{calGtc}) with the continuum limit of (\ref{bGl2}),  it can be seen, that 
$\lim\limits_{N \to \infty} \bar{\cal G}(\lambda,\lambda')=-\tilde{\cal G}_c(\lambda,\lambda'),$ 
since they satisfy the same linear integral equations{\footnote{We just note, that in the continuum limit  
$\Lambda^\pm \to \infty,$ thus $G_\Lambda(\lambda,\lambda') \to G(\lambda-\lambda').$}}. 
Thus, inserting the continuum counterpart of each function entering the final 
result{\footnote{Namely, making the replacements  $Z(h_k) \to \hat{Z}(h_k),$ $\bar{\cal G}(h_k,h_j) \to -\tilde{\cal G}_c(h_k,h_j)$ in (\ref{detcG1}).}} 
in (\ref{detcG1}),  
the continuum limit of $\lim\limits_{\eta \to 0^+} \text{det} (1-\hat{\cal G}_\eta)$ takes the form as follows: 
\begin{equation} \label{1cGc1}
\begin{split}
\lim\limits_{N \to \infty} \lim\limits_{\eta \to 0^+} \text{det} (1-\hat{\cal G}_\eta)=
 \text{det}_{ \!\!\!\! \!\!\!\!{\atop {m_H}}} \!\!\! 
\left[\delta_{jk}+\frac{\tilde{\cal G}_c(h_j,h_k)}{\hat{Z}'(h_k)} \right].
\end{split}
\end{equation}
Using (\ref{Qjkc}) this formula can be expressed by the determinant of the continuum dressed Gaudin-matrix as follows:
\begin{equation} \label{1cGc2}
\begin{split}
\lim\limits_{N \to \infty} \lim\limits_{\eta \to 0^+} \text{det} (1-\hat{\cal G}_\eta)=
 \frac{ \text{det}_{ \!\!\!\! \!\!\!\!{\atop {m_H}}} \! {\hat Q}_{jk}}{\prod\limits_{j=1}^{m_H} \hat{Z}'(h_j)}.
%\left[\delta_{jk}+\frac{\tilde{\cal G}_c(h_j,h_k)}{\hat{Z}'(h_k)} \right].
\end{split}
\end{equation}
From this formula, one can recognize, that the determinant 
$\lim\limits_{\eta \to 0^+} \text{det} (1-\hat{\cal G}_\eta)$ is 
proportional to the determinant of the Gaudin-matrix obtained directly 
from the quantization equations of the solitons.
%The importance of formula (\ref{1cGc2}) is that it accounts for the physical intuition, according to which

The last determinant one has to compute in the continuum limit is 
$\lim\limits_{\eta \to 0^+} \text{det} (1-\hat{\cal B}_\eta).$ 
The definition of the kernel $B_\eta(\lambda,\lambda)$ (\ref{Beta}) implies, that 
in the continuum limit this determinant becomes the functional determinant as follows:
\begin{equation} \label{det1pBc}
\begin{split}
\lim\limits_{N \to \infty} \lim\limits_{\eta \to 0^+} \text{det} (1-\hat{\cal B}_\eta)=
\lim\limits_{\eta \to 0^+} \text{det} (1-{\hat{\cal B}}_{\eta,c}),
\end{split}
\end{equation}
where following from (\ref{GLlim}), (\ref{calFeta}) and (\ref{calFh}) 
the kernel of the operator $\lim\limits_{N \to \infty}{\hat{\cal B}}_{\eta}={\hat{\cal B}}_{\eta,c}$ is given by:
\begin{equation} \label{Betac}
\begin{split}
{\cal B}_{\eta,c}(\lambda,\lambda')=G(\lambda-\lambda')\, \Omega_{\hat{\cal F}}^{(\eta)}(\lambda'), \qquad 
\lambda,\lambda' \in {\mathbb R},
\end{split}
\end{equation}
with 
\begin{equation} \label{OMFh}
\begin{split}
\Omega_{\hat{\cal F}}^{(\eta)}(\lambda)=\hat{\cal F}_+(\lambda+i \, \eta)+\hat{\cal F}_-(\lambda-i \, \eta)+O(\eta).
\end{split}
\end{equation}
Unfortunately, the determinant $\lim\limits_{\eta \to 0^+} \text{det} (1-\hat{\cal B}_\eta)$ remains a functional 
determinant in the continuum limit and it does not 
simplify to a finite dimensional problem as it happened in the case of the term $\lim\limits_{\eta \to 0^+} \text{det} (1-\hat{\cal G}_\eta).$ Nevertheless, it turns out, that for large values of the dimensionless volume $\ell,$ 
it can be represented by the Plemelj-series formula (\ref{Plemelj}), which turns out to be convergent:
\begin{equation} \label{det1pBplem}
\begin{split}
\lim\limits_{\eta \to 0^+}\!\!\!\text{det}(1\!+\!{\hat{\cal B}}_{\eta,c})\!=\!
\exp \!\! \left\{\sum\limits_{n=1}^{\infty} \!\! \frac{-1}{n} \!\!\! \int\limits_{-\infty}^{\infty} \!\!\! 
d\lambda_1...d\lambda_n \!\!\!\!\!\!\! \sum\limits_{\alpha_1,..,\alpha_n=\pm} \!\!\!\!\!\!\!\!
G(\lambda_1^{(\alpha_1)}\!-\!\lambda_2^{(\alpha_2)})...\!G(\lambda_n^{(\alpha_n)}\!-\!\lambda_1^{(\alpha_1)})
\! \prod\limits_{k=1}^n \! \hat{\cal F}_{\alpha_k}(\lambda_k^{(\alpha_k)})
\! \right\}\!,
\end{split}
\end{equation}
where for short, we introduced the notation: $\lambda^{(\alpha_j)}=\lambda+i \, \alpha_j \, \eta$, for all 
$j \in {\mathbb Z}_+,$ and  we exploited (\ref{OMFact}) at the derivation of this series.  
 %From (\ref{det1pBplem}) 
From the %iterative 
large volume solution of the continuum NLIE (\ref{DDVcont}), it can be shown, 
that $\lim\limits_{\ell \to \infty}\hat{\cal F}_\pm(\lambda \pm i \eta)=0.$ 
It implies, that the infinite volume limit of the functional determinant term  
becomes one:
\begin{equation} \label{contegy}
\begin{split}
\lim\limits_{\ell \to \infty}\lim\limits_{\eta \to 0^+}\!\!\!\text{det}(1\!+\!{\hat{\cal B}}_{\eta,c})=1.
\end{split}
\end{equation}
%where $\ell$ denotes the volume of the continuum quantum field theory in the units of the soliton (fermion) 
%mass.

\section{Computing the prefactor $v_0$} \label{V0sect}

The norm of a Bethe-wave function is given by the formula (\ref{norm1}). In the previous section 
we computed the determinant part $\text{det}\Phi$ 
%of (\ref{norm1}) 
and in this section we compute the 
product part $v_0.$ 
%of this formula (\ref{v0}). 
This term is a ratio of two double products.
The logarithm of these products become sums and the sums can be 
evaluated with the help of the summation formula (\ref{Sfinf}). The numerator and the denominator in  
(\ref{v0}) requires a bit different treatment. This is why we discuss their computation separately. 

\subsection{Computing the numerator} \label{NUMsubsect}

According to (\ref{v0}) the numerator of $v_0$  takes the form as follows:
\begin{equation} \label{vn1}
\begin{split}
v_n=\prod\limits_{j,k=1}^m \, \sinh(\lambda_k-\lambda_j-i \, \gamma).
\end{split}
\end{equation}
Its logarithm is a double sum:
\begin{equation} \label{lvn1}
\begin{split}
\ln v_n=\sum\limits_{j,k=1}^m \, \ln \sinh(\lambda_k-\lambda_j-i \, \gamma).
\end{split}
\end{equation}
The computation of $v_n$ consists of two steps.
First, for real $\lambda $ one computes the sum: 
\begin{equation} \label{S1k}
\begin{split}
S_1(\lambda)=\sum\limits_{j=1}^m \ln \sinh(\lambda-\lambda_j-i \, \gamma).
\end{split}
\end{equation}
Then the sum: $\sum\limits_{k=1}^m S_1(\lambda_k)$ should be evaluated to get the final result for $\ln v_n.$ 

The function $\ln \sinh(\lambda-z-i \, \gamma)$ in the summand is analytic in the strip $-\gamma<\text{Im} z <\pi-\gamma,$ 
if $\lambda \in {\mathbb R}.$ This is why the summation formula (\ref{Sfinf}) can be directly applied to (\ref{S1k}):
\begin{equation} \label{S1lam}
\begin{split}
S_1(\lambda)&=-\sum\limits_{j=1}^{m_H} \ln \sinh(\lambda-h_j-i \, \gamma)+\int\limits_{-\infty}^{\infty} \! \frac{d\lambda'}{2 \pi} \,
\ln \sinh(\lambda-\lambda'-i \, \gamma) \, Z'(\lambda)- \\
&-\sum\limits_{\alpha=\pm} \int\limits_{-\infty}^{\infty} \! \frac{d\lambda'}{2 \pi} \, \ln\sinh(\lambda-\lambda'-i \, \alpha \, \eta-i \, \gamma) \,
Z'(\lambda'+i \, \alpha \, \eta) \, {\cal F}_\alpha(\lambda'+i \,\alpha \, \eta), \qquad \lambda \in {\mathbb R}.
\end{split}
\end{equation}
The value of the contour deformation parameter is restricted by the analyticity range of the $\ln \sinh$ function and the location of poles 
of the functions $Z'(\lambda) {\cal F}_{\pm}(\lambda).$ 
From the definition (\ref{countfv}), it follows that these latter functions are free of poles in the 
range $0<|\text{Im} \lambda|<\tfrac{\gamma}{2}.$ Thus the allowed values of $\eta$ in (\ref{S1lam}) are restricted by the inequality:
\begin{equation} \label{etan1}
\begin{split}
0<\eta<\text{min}(\tfrac{\gamma}{2},\pi-\gamma).
\end{split}
\end{equation}
We just note that (\ref{etan1}) is obtained by assuming that the absolute values of the imaginary parts 
of the contour 
in the upper and lower half planes are equal. Without this symmetry two contour deformation parameters 
could be introduced, such that 
each satisfy an inequality similar to (\ref{etan1}).

To get the final formula for the numerator, the  sum $\sum\limits_{k=1}^m S_1(\lambda_k)$ should also be computed. 
Formula (\ref{S1lam}) implies, that this requires the transformation of the following sums into integral expressions:
\begin{equation} \label{Pgpm}
\begin{split}
P_\gamma(\lambda')&=\sum\limits_{k=1}^m \ln\sinh(\lambda_k-\lambda'-i \, \gamma), \qquad \qquad \quad \lambda'\in{\mathbb R}, \\
P_{\pm}(\lambda')&=\sum\limits_{k=1}^m \ln\sinh(\lambda_k-(\lambda'\pm i \, \eta)-i\, \gamma), \qquad \lambda'\in{\mathbb R}.
\end{split}
\end{equation}
With the help of the summation formula (\ref{Sfinf}) these sums can be expressed in terms of the holes and 
the counting function in a straightforward manner. They take the forms as follows:
\begin{equation} \label{Pg}
\begin{split}
P_\gamma(\lambda')&=-\sum\limits_{s=1}^{m_H} \ln\sinh(h_s-\lambda'-i \, \gamma)+\int\limits_{-\infty}^{\infty} \!
\frac{d\lambda}{2 \pi} \,\, \ln\sinh(\lambda-\lambda'-i\, \gamma) \, Z'(\lambda)- \\
&-\sum\limits_{\alpha=\pm} \int\limits_{-\infty}^{\infty} \frac{d\lambda}{2 \pi} \ln\sinh(\lambda-\lambda'+i \, \alpha \, \eta'-i\, \gamma) \,
Z'(\lambda+i  \, \alpha \, \eta') \, {\cal F}_\alpha(\lambda+i \, \alpha \, \eta'),
\end{split}
\end{equation}
\begin{equation} \label{Ppm}
\begin{split}
P_\pm(\lambda')&=-\sum\limits_{s=1}^{m_H} \ln\sinh(h_s-(\lambda'\pm i \, \eta)-i \, \gamma)+\int\limits_{-\infty}^{\infty} \!
\frac{d\lambda}{2 \pi} \,\, \ln\sinh(\lambda-(\lambda'\pm i \, \eta)-i\, \gamma) \, Z'(\lambda)- \\
&-\sum\limits_{\alpha=\pm} \int\limits_{-\infty}^{\infty} \frac{d\lambda}{2 \pi} \ln\sinh(\lambda-(\lambda'\pm i \, \eta)+i \, \alpha \, \eta_\alpha-i\, \gamma) \,
Z'(\lambda+i  \, \alpha \, \eta_\alpha) \, {\cal F}_\alpha(\lambda+i \, \alpha \, \eta_\alpha),
\end{split}
\end{equation}
where the range for the contour deformation parameters are restricted by the inequalities as follows: 
\begin{equation} \label{etagpm}
\begin{split}
0<\eta'<\text{min}(\tfrac{\gamma}{2},\pi-\gamma), \qquad 0<\eta_\pm<\text{min}(\tfrac{\gamma}{2},\pi-\gamma \mp\eta).
\end{split}
\end{equation}
Using the expressions (\ref{Pg}) and (\ref{Ppm}) together with (\ref{S1lam}), $\ln v_n$ can be written as follows:
\begin{equation} \label{lnvnfinal}
\begin{split}
\ln v_n=-\sum\limits_{j=1}^{m_H} P_\gamma(h_j)\!+\!\!\int\limits_{-\infty}^{\infty} \!\! \frac{d\lambda'}{2 \pi} P_\gamma(\lambda') \, Z'(\lambda')-\!\!
\sum\limits_{\alpha=\pm} \int\limits_{-\infty}^{\infty} \!\! \frac{d\lambda'}{2 \pi} P_\alpha(\lambda') \, Z'(\lambda'+i \, \alpha \, \eta) \, {\cal F}_{\alpha}(\lambda'+i \, \alpha \, \eta),
\end{split}
\end{equation}
with $\eta$ being restricted by (\ref{etan1}).

We note that the formula (\ref{lnvnfinal}) could be discussed in the large $N$ limit. Nevertheless, we refrain from 
the careful discussion of the large $N$ limit of this formula, because the result would be quite complicated and 
in many cases useless.  
The reason is that the norm of a Bethe-eigenstate alone is not a physically interesting quantity. 
The interesting physical quantities are the form-factors of the local operators of the theory and  
in many cases the multiplicative factors like $v_n$ simply cancel from their final formula, making 
unnecessary to determine such products in the continuum limit.

%It becomes interesting, when it arises in the formula for the form-factor of some local operator of the theory (See %(\ref{FO1}) in the introduction). In many cases in the formulas for the form-factors multiplicative factors 
%similar to $v_n$ arise

%discussing the continuum limit, since in many 
%cases these prefactors cancel from an interesting physical quantity like a non-diagonal form-factor. 

\subsection{Computing the denominator} \label{DENOMsect}

\begin{figure}[htb]
\begin{flushleft}
%\vskip 10mm
\hskip 15mm
\leavevmode
\epsfxsize=120mm
\epsfbox{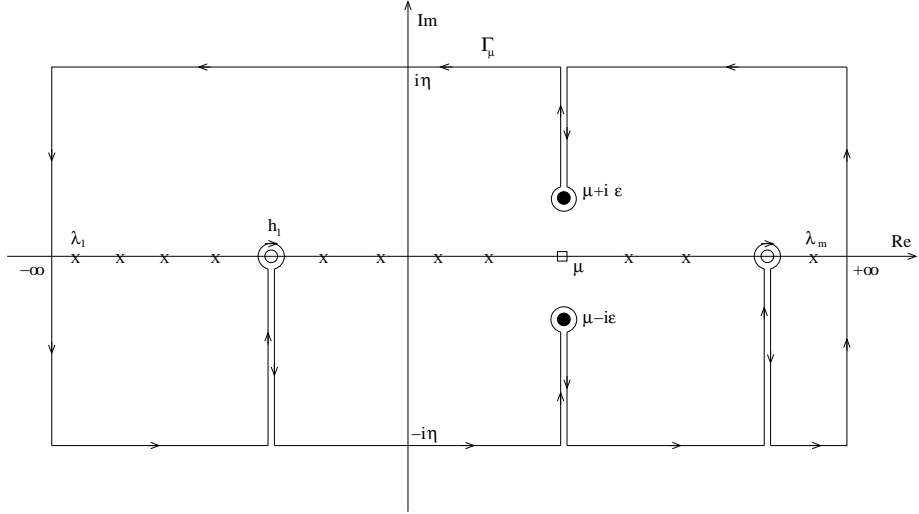}
%\vskip 10mm
\end{flushleft}
\caption{{\footnotesize
The figure represents the closed curve $\Gamma_\mu.$ Crosses and circles represent the Bethe-roots and holes respectively. The tiny rectangle denotes the value of $\mu$  and 
black filled circles denote the poles at $\mu \pm i \, \epsilon.$
}}
\label{figGmu}
\end{figure}

The computation of the denominator of $v_0$ (\ref{v0}) is less straightforward than that of the numerator. 
It is given by the product as follows:
\begin{equation} \label{vd}
\begin{split}
v_d=\prod\limits_{j>k}^m \sinh(\lambda_j-\lambda_k) \, \sinh(\lambda_k-\lambda_j).
\end{split}
\end{equation}
The main difficulty comes from the fact, that the $j=k$ cases must be omitted from the ranges of 
the indexes $j$ and $k.$
Thus, to rephrase the logarithm of $v_d$ as an integral expression, first one has to make some simple transformations on (\ref{vd}).
Since $\sinh(\lambda_j-\lambda_k)=-\sinh(\lambda_k-\lambda_j),$ (\ref{vd}) can be written as follows:
\begin{equation} \label{vd1}
\begin{split}
v_d=(-1)^{\tfrac{m(m-1)}{2}} \, {\cal N}_0, \qquad {\cal N}_0=\prod\limits_{j>k}^m \sinh(\lambda_j-\lambda_k)^2.
\end{split}
\end{equation}
We treat the case, when all Bethe-roots are real, then it follows, that ${\cal N}_0$ is real and positive. 
The quantity, we will compute is the logarithm of ${\cal N}_0^2.$ In terms of this quantity $v_d$ can be given as follows:
\begin{equation} \label{vd3}
\begin{split}
v_d=(-1)^{\tfrac{m(m-1)}{2}} \, e^{\tfrac{1}{2}\ln {\cal N}_0^2}, 
\end{split}
\end{equation}
such that ${\cal N}_0^2$ is given by:
\begin{equation} \label{calN02}
\begin{split}
{\cal N}_0^2=\prod\limits_{j,k=1 \atop j\neq k}^m \sinh(\lambda_j-\lambda_k)^2.
\end{split}
\end{equation}
For the actual computations it is worth to introduce the functions:
\begin{equation} \label{lnSpm}
\begin{split}
\ln S_+(x)=\ln(i \, \sinh(x))-\ln i, \qquad \ln S_-(x)=\ln(\tfrac{1}{i} \, \sinh(x))-\ln \tfrac{1}{i}.
\end{split}
\end{equation}
They have the important analytical properties as follows: 
\begin{itemize}
\item $\ln S_+(x)$ is analytical in the range: $-\pi<\text{Im} \, x<0,$
\item $\ln S_+(x)$ is analytical in the range: $0<\text{Im} \, x<\tfrac{\pi}{2},$ apart from a 
vertical logarithmic cut with a jump given by the formula:
\begin{equation} \label{jumpp}
\begin{split}
\lim\limits_{\epsilon \to 0^+}\left\{\ln S_+(i \, t+  \epsilon)-\ln S_+(i \, t- \epsilon)\right\}=2 \pi \, i, \qquad 0<t<\tfrac{\pi}{2},
\end{split}
\end{equation}
\item $\ln S_-(x)$ is analytical in the range: $0<\text{Im} \, x<\pi,$
\item $\ln S_-(x)$ is analytical in the range: $-\tfrac{\pi}{2}<\text{Im} \, x<0,$ apart from a 
vertical logarithmic cut with a jump given by the formula:
\begin{equation} \label{jumpm}
\begin{split}
\lim\limits_{\epsilon \to 0^+}\left\{\ln S_-(i \, t+\epsilon)-\ln S_-(i \, t- \epsilon)\right\}=-2 \pi \, i, \qquad -\tfrac{\pi}{2}<t<0.
\end{split}
\end{equation}
\end{itemize}
The use of these functions is that they provide a regularization for the function $\ln\sinh^2$ by the following formula:
\begin{equation} \label{lnSpmreg}
\begin{split}
\lim\limits_{\epsilon \to 0^+}\left\{\ln S_+(x-i \, \epsilon)+\ln S_-(x+i \, \epsilon)\right\}=\ln \sinh^2(x), \qquad \forall x \in {\mathbb R.}
\end{split}
\end{equation}
Now, it is worth to define the function:
\begin{equation} \label{Fe}
\begin{split}
F_\epsilon(\mu,\lambda)=\ln S_+(\mu-\lambda-i \, \epsilon)+\ln S_-(\mu-\lambda+i \, \epsilon).
\end{split}
\end{equation}
From the properties of $\ln S_\pm(x)$ it follows, that $F_\epsilon(\mu,\lambda)$ is regular in an $\epsilon$ wide neighborhood of the real axis.
Then the logarithm of ${\cal N}_0^2$ in (\ref{calN02}) can be computed as follows:
\begin{equation} \label{lcN02}
\begin{split}
\ln {\cal N}_0^2=\lim\limits_{\epsilon \to 0^+} \left( \sum\limits_{j,k=1}^m F_\epsilon(\lambda_k,\lambda_j)-\sum\limits_{j=1}^m F_\epsilon(0) \right).
\end{split}
\end{equation}
Using the small argument series for $\ln S_\pm(x),$ (\ref{lcN02}) can be written as follows:
\begin{equation} \label{lcN02a}
\begin{split}
\ln {\cal N}_0^2=\lim\limits_{\epsilon \to 0^+} \left( \Sigma_\epsilon-
m \, (\ln(i \, \epsilon)+\ln(-i \, \epsilon))\right),
\end{split}
\end{equation}
where we introduced the notation:
\begin{equation} \label{Seps1}
\begin{split}
\Sigma_\epsilon=\sum\limits_{j,k=1}^m F_\epsilon(\lambda_k,\lambda_j).
\end{split}
\end{equation}
In the sequel we compute $\Sigma_\epsilon$ upto of order one in the $\epsilon \to 0^+$ limit.
As a first step the sum:
\begin{equation} \label{Sig11}
\begin{split}
\Sigma_1(\mu)=\sum\limits_{j=1}^m F_\epsilon(\mu,\lambda_j), \qquad \mu \in {\mathbb R,}
\end{split}
\end{equation}
should be computed for real values of $\mu.$ To avoid complications coming from the discontinuities of 
$F_\epsilon(\mu,\lambda),$ first we compute the derivative:
\begin{equation} \label{DSig1}
\begin{split}
\Sigma_1'(\mu)=\sum\limits_{j=1}^m F_\epsilon'(\mu,\lambda_j), 
\qquad F_\epsilon'(\mu,\lambda_j)=\partial_{\mu} F_\epsilon(\mu,\lambda_j).
\end{split}
\end{equation}
This can be done by a computation similar to that presented in section \ref{SUMsect} at the derivation 
of the summation formulas. Namely, the sum can be rewritten as a contour integral along 
the contour $\Gamma_\mu$ depicted in figure \ref{figGmu}.
%, such that the curve can run to infinity in the horizontal direction (I.e. $\Lambda^\pm \to\infty$).  
%%%%%%%%%%%%%%%%%%%%%%%%
%%%%%%%%%%%%%%%%%%%%%%%%% 
Then, with the help of the residue theorem and the following residues: 
\begin{equation} \label{residues1}
\begin{split}
\text{Res}_{\lambda=\mu \pm i \, \epsilon} F_\epsilon'(\mu,\lambda)=-1, \qquad 
\text{Res}_{\lambda=\lambda_j,h_j} Z'(\lambda){\cal F}_+(\lambda)=-i,
\end{split}
\end{equation}
one ends up with the following result:
\begin{equation} \label{DSig2}
\begin{split}
\Sigma^{\prime}_1(\mu)&=\int\limits_{-\infty}^{\infty}\! \frac{d\lambda}{2\pi} \,
 F_\epsilon'(\mu,\lambda-i \, \eta) \, Z'(\lambda-i \, \eta) -\sum\limits_{\alpha=\pm} \int\limits_{-\infty}^{\infty}
\frac{d\lambda}{2 \pi} F_\epsilon'(\mu,\lambda^{(\alpha)}) \, Z'(\lambda^{(\alpha)}) \, 
{\cal F}_\alpha(\lambda^{(\alpha)}) \\
&-\sum\limits_{j=1}^{m_H} F_\epsilon'(\mu,h_j)
+\frac{d}{d\mu}\left[\ln \left( 1+(-1)^\delta e^{i \, Z(\mu+i\, \epsilon)}\right)+
\ln \left( 1+(-1)^\delta e^{i \, Z(\mu-i\, \epsilon)}\right) \right],
\end{split}
\end{equation}
where $\lambda^{(\alpha)}=\lambda+i \, \alpha \, \eta,$ and we exploited, that 
\begin{equation} \label{exploit1}
\begin{split}
Z'(\lambda){\cal F}_+(\lambda)=-i \frac{d}{d\lambda}\ln \left( 1+(-1)^\delta e^{i \, Z(\lambda)}\right), \qquad
\lim\limits_{\lambda \to \pm \infty} F_\epsilon'(\mu,\lambda) \, Z'(\lambda){\cal F}_+(\lambda)=0.
\end{split}
\end{equation}
The next step is to integrate (\ref{DSig2}) %. We will do that 
upto $O(\epsilon)$ corrections. One can recognize, that 
the integral terms of (\ref{DSig2}) are regular in the $\epsilon \to 0^+$ limit. 
This is why we take the $\epsilon \to 0^+$ limit 
in these terms when integrating (\ref{DSig2}). 
Consider the function: 
\begin{equation} \label{Re1}
\begin{split}
R_\epsilon(\mu)&=\int\limits_{-\infty}^{\infty}\! \frac{d\lambda}{2 \pi} \, 2  
 \ln S_-(\mu-\lambda^{(-)}) \, Z'(\lambda^{(-)}) -\sum\limits_{\alpha=\pm} \int\limits_{-\infty}^{\infty}
\frac{d\lambda}{2 \pi} \, 2 \ln S_\alpha(\mu-\lambda^{(\alpha)}) \, Z'(\lambda^{(\alpha)}) \, 
{\cal F}_\alpha(\lambda^{(\alpha)}) \\
&-\sum\limits_{j=1}^{m_H} F_\epsilon(\mu,h_j)
+\sum\limits_{\alpha=\pm} \ln \left( 1+(-1)^\delta e^{i \, Z(\mu+i\, \alpha \, \epsilon)}\right)-2 \,
\ln \left( 1+(-1)^\delta e^{i \, Z(+\infty)}\right).
\end{split}
\end{equation}
By differentiating (\ref{Re1}) with respect to $\mu,$ 
and using the identities following from (\ref{Sfinf}) as follows:
\begin{equation} \label{ids1}
\begin{split}
m&=-m_H+\int\limits_{-\infty}^{\infty} \frac{d\lambda}{2 \pi} Z'(\lambda-i\, \eta)-
\sum\limits_{\alpha=\pm} \int\limits_{-\infty}^{\infty} \frac{d\lambda}{2 \pi} Z'(\lambda+i \, \alpha \, \eta)\,
{\cal F}_{\alpha}(\lambda+i \, \alpha \, \eta), \\
\sum\limits_{j=1}^m \lambda_j&=-\!\!\sum\limits_{j=1}^{m_H} \lambda_j+\!\!\!
\int\limits_{-\infty}^{\infty} \!\! \frac{d\lambda}{2 \pi} \! (\lambda-i\, \eta) Z'(\lambda-i\, \eta)-
\!\!\! \sum\limits_{\alpha=\pm} \!\! \int\limits_{-\infty}^{\infty} \!\! \frac{d\lambda}{2 \pi} (\lambda+i \, \alpha \, \eta)
Z'(\lambda+i \, \alpha \, \eta)\,
{\cal F}_{\alpha}(\lambda+i \, \alpha \, \eta), 
\end{split}
\end{equation}
it can be shown, that:
\begin{equation} \label{SRe}
\begin{split}
\Sigma_1'(\mu)=\frac{d}{d \mu} R_\epsilon(\mu)+O(\epsilon), \qquad 
\lim\limits_{\mu \to \infty} \left( \Sigma_1(\mu) -R_\epsilon(\mu)\right)=0. 
\end{split}
\end{equation}
It follows, that:
\begin{equation} \label{SRe1}
\begin{split}
\Sigma_1(\mu)= R_\epsilon(\mu)+O(\epsilon).
\end{split}
\end{equation}
%In the sequel we  will not indicate the $O(\epsilon)$ terms, since the computations are interesting upto 
%the constant in $\epsilon$ order. 

Then, one has to take the second sum in (\ref{Seps1}). 
From (\ref{SRe1}) it follows, that the following sum should be computed:
\begin{equation} \label{Seps2}
\begin{split}
\Sigma_\epsilon=\sum\limits_{k=1}^m R_\epsilon(\lambda_k)+O(\epsilon).
\end{split}
\end{equation}
Formula (\ref{Re1}) for $R_\epsilon(\mu)$ implies, that this requires the 
computation of the following sums:
\begin{eqnarray} \label{O1j}
{\cal O}_1^{(j)}&=&\sum\limits_{k=1}^m F_\epsilon(\lambda_k,h_j), \\ 
{\cal O}_\alpha(\lambda)&=&\sum\limits_{k=1}^m 2 \, \ln S_{\alpha}(\lambda_k-(\lambda+i \, \alpha \, \eta)), \qquad \alpha=\pm, \quad \lambda \in {\mathbb R}, 
\label{Oalfa} \\ 
{\cal O}^{(\epsilon)}_L&=&\sum\limits_{k=1}^m \left\{ \ln \left( 1+(-1)^\delta e^{i \, Z(\lambda_k-i\, \epsilon)}\right)+\ln \left( 1+(-1)^\delta e^{i \, Z(\lambda_k+i\, \epsilon)}\right)   \right\}.
\label{OLe}
\end{eqnarray}
A straightforward computation similar to the derivation of the summation formulas in section \ref{SUMsect}, 
leads to the following result for ${\cal O}_\pm(\lambda):$
\begin{equation} \label{Oalfa1}
\begin{split}
{\cal O}_\alpha(\lambda)&=-2 \sum\limits_{j=1}^{m_H} \ln S_\alpha(h_j-\lambda-i \, \alpha \,\eta)\!+\!\!\! \int\limits_{-\infty}^{\infty} \frac{d\lambda'}{\pi} \,
\ln S_\alpha(\lambda'-\lambda-i \, \alpha \, \eta)\, Z'(\lambda')-\\
&-\!\! \sum\limits_{\beta=\pm} \!\! \int\limits_{-\infty}^{\infty} \frac{d\lambda'}{\pi} \,\ln S_{\beta}(\lambda'-\lambda+i\, \beta \, \eta'-i \, \alpha \, \eta )\,
Z'(\lambda'+i \, \beta \, \eta') \, {\cal F}_\beta(\lambda'+i \, \beta \, \eta'), \quad \alpha=\pm,
\end{split}
\end{equation}
where the contour deformation parameter $\eta'$ is chosen to avoid the branch cuts of $\ln S_\pm(\lambda):$
\begin{equation} \label{eta'}
\begin{split}
0<\eta'<\text{min}(\eta,\pi-\eta), \qquad 0<\eta<\text{min}(\tfrac{\gamma}{2},\pi-\gamma).
\end{split}
\end{equation}
Next we compute the sum ${\cal O}_L^{(\epsilon)}$ defined in (\ref{OLe}) in the $\epsilon \to 0^+$ limit.  
As a consequence of the Bethe-equations (\ref{BAEZ}) this sum is divergent in this limit, since: 
\begin{equation} \label{BAEZ1}
\begin{split}
1+(-1)^\delta \, e^{i \, Z(\lambda_k)}=0, \qquad k=1,...,m.
\end{split}
\end{equation}
Close to the Bethe-root $\lambda_k,$ the behavior of $\ln\left( 1+(-1)^\delta \, e^{i \, Z(\lambda)}\right)$ is given by the formula:
\begin{equation} \label{zllk}
\begin{split}
\ln\left( 1+(-1)^\delta \, e^{i \, Z(\lambda)}\right) \stackrel{\lambda \sim \lambda_k}{=} \ln \left[ c_k (\lambda-\lambda_k)+O(\lambda-\lambda_k)^2 \right],
\end{split}
\end{equation}
where
\begin{equation} \label{ck}
\begin{split}
c_k=\frac{d}{d\lambda} \ln\left( 1+(-1)^\delta \, e^{i \, Z(\lambda)}\right)\bigg|_{\lambda=\lambda_k}=(-1)^\delta \, e^{i \, Z(\lambda_k)} \, i\, Z'(\lambda_k)=-i\, Z'(\lambda_k).
\end{split}
\end{equation}
Using (\ref{zllk}) and (\ref{ck}) one obtains the result as follows:
\begin{equation} \label{OLe1}
\begin{split}
{\cal O}_L^{(\epsilon)}={\cal O}^{reg}_L+m \, (\ln(i \, \epsilon)+\ln(-i \, \epsilon))+O(\epsilon),
\end{split}
\end{equation}
where ${\cal O}^{reg}_L$ is the regular part:
\begin{equation} \label{OLreg}
\begin{split}
{\cal O}^{reg}_L=2 \sum\limits_{k=1}^m \, \ln[-i \, Z'(\lambda_k)].
\end{split}
\end{equation}
Formula (\ref{OLe1}) accounts for the divergent part of $\Sigma_\epsilon.$ The divergent term arising in (\ref{OLe1}) 
cancels from the physically interesting quantity $\ln {\cal N}_0^2$ due to (\ref{lcN02a}).
The sum in (\ref{OLreg}) could be rephrased as an integral expression with the help of the formulas of section 
\ref{SUMsect}, but we 
do not carry out this transformation, since in (\ref{detFi1}) a similar term arises, which cancel the contribution of 
$O_L^{reg}$ from the final formula for the norm of the Bethe-state. We will discuss this cancellation in more detail 
in the summary of the paper.

The last sum one has to compute is ${\cal O}_1^{(j)}$ defined in (\ref{O1j}). 
%Here we rephrase it by introducing some new notations:
To emphasize, that $F_\epsilon(\lambda,h_j)$ in (\ref{O1j}) is considered as the function of a  
single variable $\lambda,$ we rewrite (\ref{O1j}) in the form as follows:
\begin{equation} \label{O1j1}
\begin{split}
{\cal O}_1^{(j)}=\sum\limits_{k=1}^m \ln S_\epsilon^{(j)}(\lambda_k), \qquad  
\ln S_\epsilon^{(j)}(\lambda)=\ln S_+(\lambda-h_j-i \, \epsilon)+\ln S_-(\lambda-h_j+i \, \epsilon).
\end{split}
\end{equation}
The function $\ln S_\epsilon^{(j)}(\lambda)$ is analytic in an $\epsilon$ wide strip around the real axis, and 
as a consequence of (\ref{jumpp}) and (\ref{jumpm}) it has vertical discontinuities starting from the points $h_j \pm i \,\epsilon.$
Using the residue theorem, ${\cal O}_1^{(j)}$  can be written as a contour integral along the curve $\Gamma_j$ depicted in figure \ref{figGj}:
%%%%%%%%%%%%%%%%%%%%%%%%
 \begin{figure}[htb]
\begin{flushleft}
%\vskip 10mm
\hskip 15mm
\leavevmode
\epsfxsize=120mm
\epsfbox{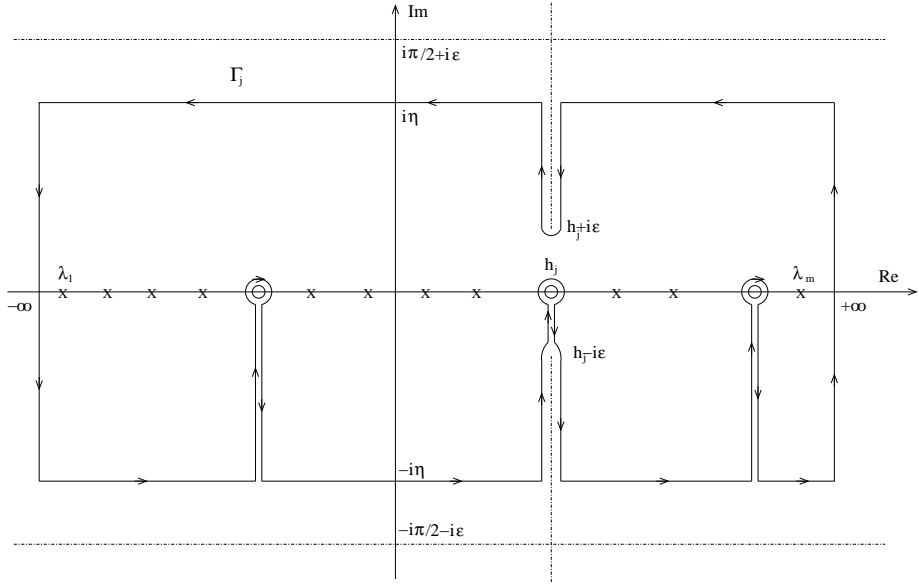}
%\vskip 10mm
\end{flushleft}
\caption{{\footnotesize
The figure represents the closed curve $\Gamma_j$ in (\ref{O1j2}). Crosses and circles represent the Bethe-roots and holes respectively. 
Dashed lines represent the discontinuities of the integrand.
%The distance of the horizontal lines from the real axis is indicated at the right hand side of the curve ($\pm i \eta$).
}}
\label{figGj}
\end{figure} 
%%%%%%%%%%%%%%%%%%%%%%%%% 
\begin{equation} \label{O1j2}
\begin{split}
{\cal O}_1^{(j)}&=\oint\limits_{\Gamma_j} \frac{d\lambda}{2 \pi} \, \ln S_\epsilon^{(j)}(\lambda)\, Z'(\lambda) \, {\cal F}_+(\lambda)= \\
&=-\sum\limits_{s=1}^{m_H} \ln S_\epsilon^{(j)}(h_s)+Q_{cut}^{(j)}
-\sum\limits_{\alpha=\pm} \alpha \int\limits_{h_j} \!\!\!\!\!\!- \frac{d\lambda}{2 \pi} 
\ln S_\epsilon^{(j)}(\lambda^{(\alpha)} ) \, Z'(\lambda^{(\alpha)}) \, {\cal F}_+(\lambda^{(\alpha)}),
\end{split}
\end{equation}
where for $\alpha=\pm:$ $\lambda^{(\alpha)}=\lambda+i \, \alpha \, \eta,$ $0<\eta<\tfrac{\gamma}{2},$ and the symbol $\int\limits_{h_j} \!\!\!\!\!\!-$ 
denotes a principal value integration. This principal value integration is not introduced, because the integrand is singular at the point $h_j,$ 
but in order to indicate that the integrand is discontinuous at the point $h_j.$ The symbol $Q_{cut}^{(j)}$ stands for that part of the contour integral, 
which belongs to encircling the vertical discontinuities. From (\ref{jumpp}) and (\ref{jumpm}) it can be computed explicitly:  
\begin{equation} \label{Qjcut1}
\begin{split}
Q_{cut}^{(j)}=\sum\limits_{\alpha=\pm} \left(\ln(1+(-1)^\delta e^{i \, Z(h_j+i \, \alpha \, \epsilon)})-\ln(1+(-1)^\delta e^{i \, Z(h_j+i \, \alpha \, \eta)})\right).
\end{split}
\end{equation}
As a consequence of (\ref{BAEH}) $1+(-1)^\delta e^{i \, Z(h_j)}=0.$ Thus after an analog computation to that 
of (\ref{BAEZ1})-(\ref{OLreg}), one obtains for small $\epsilon$ 
the result as follows:
\begin{equation} \label{Qjcut2}
\begin{split}
Q_{cut}^{(j)}=\ln\left[ -i \, Z'(h_j)\right]+\ln(i \, \epsilon)+\ln(-i \, \epsilon)-\sum\limits_{\alpha=\pm} \ln(1+(-1)^\delta e^{i \, Z(h_j+i \, \alpha \, \eta)})+
O(\epsilon).
\end{split}
\end{equation}
It implies, that $Q_{cut}^{(j)}$ is divergent in the $\epsilon \to 0^+$ limit. On the other hand, from (\ref{O1j1}), (\ref{lnSpm}) it can be shown, that
the sum $-\sum\limits_{s=1}^{m_H} \ln S_\epsilon^{(j)}(h_s)$ in (\ref{O1j2}) is also divergent in the $\epsilon \to 0^+$ limit:
\begin{equation} \label{sumhj}
\begin{split}
-\sum\limits_{s=1}^{m_H} \ln S_\epsilon^{(j)}(h_s)=-\sum\limits_{s=1 \atop s\neq j}^{m_H} \ln \sinh^2(h_s-h_j)-\ln(i \, \epsilon)-\ln(-i\, \epsilon)+O(\epsilon).
\end{split}
\end{equation}
Adding (\ref{Qjcut2}) and (\ref{sumhj}) together as required by (\ref{O1j2}), it can be seen, that the divergent term cancels from the relevant sum ${\cal O}_1^{(j)}.$
After making the replacement ${\cal F}_+(\lambda-i \, \eta) \to 1-{\cal F}_-(\lambda-i \, \eta)$ in (\ref{O1j2}), 
${\cal O}_1^{(j)}$ can be written as follows:
\begin{equation} \label{O1j3}
\begin{split}
{\cal O}_1^{(j)}&=\ln\left[ -i \, Z'(h_j)\right]-\sum\limits_{s=1 \atop s\neq j}^{m_H} \ln \sinh^2(h_s-h_j)
-\sum\limits_{\alpha=\pm} \ln(1+(-1)^\delta e^{i \, Z(h_j+i \, \alpha \, \eta)})- \\
&-\sum\limits_{\alpha=\pm}  \int\limits_{h_j} \!\!\!\!\!\!- \frac{d\lambda}{2 \pi} 
\ln S_\epsilon^{(j)}(\lambda^{(\alpha)} ) \, Z'(\lambda^{(\alpha)}) \, {\cal F}_\alpha(\lambda^{(\alpha)})+
\int\limits_{h_j} \!\!\!\!\!\!- \frac{d\lambda}{2 \pi} \ln S_\epsilon^{(j)}(\lambda-i\, \eta ) \, Z'(\lambda-i \, \eta)+O(\epsilon).
\end{split}
\end{equation}
The last integral can be shifted to the real axis by appropriate deformation of the integration contour:
\begin{equation} \label{intZv}
\begin{split}
\int\limits_{h_j} \!\!\!\!\!\!- \frac{d\lambda}{2 \pi} \ln S_\epsilon^{(j)}(\lambda-i\, \eta ) \, Z'(\lambda-i \, \eta)\!=\!\!\!\!
\int\limits_{-\infty}^\infty \!\! \frac{d\lambda}{2 \pi} \ln \sinh^2(\lambda-h_j) \, Z'(\lambda)\!-\!i \, Z(h_j)\!+\!i \,Z(h_j-i \, \eta)+O(\epsilon),
\end{split}
\end{equation}
where the non-integral terms come from the contributions of the vertical discontinuity of the integrand and we took the $\epsilon \to 0^+$ 
limit. 
With the help of (\ref{BAEH}) one can derive the identity:
\begin{equation} \label{idhole}
\begin{split}
-i  Z(h_j)\!\!+\!\!i Z(h_j-i\, \eta)\!\!-\!\!\ln\left( 1+(-1)^\delta e^{i \, Z(h_j-i \, \eta)}\right)\!\!=\!\!-\ln\left( 1+(-1)^\delta e^{-i \, Z(h_j-i \, \eta)}\right)\!\!
-\!\!2 \pi i \, n_j\!\!-\!\!i \pi,  
\end{split}
\end{equation}
with $n_j=I_{h_j}-\tfrac{1}{2} \in {\mathbb Z}.$
Inserting (\ref{intZv}) and (\ref{idhole}) into (\ref{O1j3}) one ends up with the result:
\begin{equation} \label{O1j4}
\begin{split}
{\cal O}^{(j)}_1=2 \ln \left[-i \, Z'(h_j) \right]+\hat{\cal O}^{(j)}_1-i \, \pi+O(\epsilon) \,
\, \,  \text{mod} \,\,  2 \pi \,i,
\end{split}
\end{equation}
where 
\begin{equation} \label{hatO1j}
\begin{split}
\hat{\cal O}_1^{(j)}&=-\sum\limits_{s=1 \atop s\neq j}^{m_H} \ln \sinh^2(h_s-h_j)
-\sum\limits_{\alpha=\pm} \ln\left(1+(-1)^\delta e^{\alpha \, i \, Z(h_j+i \, \alpha \, \eta)}\right)- \\
&-\sum\limits_{\alpha=\pm}  \int\limits_{h_j} \!\!\!\!\!\!- \frac{d\lambda}{2 \pi} 
\ln \sinh^2(\lambda^{(\alpha)} -h_j) \, Z'(\lambda^{(\alpha)}) \, {\cal F}_\alpha(\lambda^{(\alpha)})+
\int\limits_{-\infty}^{\infty}  \frac{d\lambda}{2 \pi} \ln \sinh^2(\lambda-h_j) \, Z'(\lambda).
\end{split}
\end{equation}
We described the result only modulo $2 \pi \, i$ terms, since the norm of the Bethe-eigenstates 
contain only the exponential of this quantity. 
Using the definitions (\ref{O1j})-(\ref{OLe}) together with (\ref{Seps2}) and (\ref{Re1}) one obtains for $\Sigma_\epsilon$ the following result:
\begin{equation} \label{Sigmae1}
\begin{split}
\Sigma_\epsilon&=-\sum\limits_{j=1}^{m_H} \, {\cal O}_1^{(j)}+{\cal O}^{(\epsilon)}_L\!\!+\!\! 
\int\limits_{-\infty}^{\infty} \frac{d \lambda}{2 \pi} \, {\cal O}_-(\lambda-i \, \eta) \, Z'(\lambda-i \, \eta)-\\
&-\sum\limits_{\alpha=\pm} \int\limits_{-\infty}^{\infty} \frac{d \lambda}{2 \pi} {\cal O}_\alpha(\lambda+i \, \alpha \, \eta) \,
Z'(\lambda+i \, \alpha \, \eta) \, {\cal F}_\alpha(\lambda+i \, \alpha \, \eta)
-2 \,m \, \ln \left(1+(-1)^{\delta} e^{i \, Z(+\infty)} \right),
\end{split}
\end{equation}
where ${\cal O}_1^{(j)},$ ${\cal O}_\pm(\lambda)$ and ${\cal O}^{(\epsilon)}_L$ are given by the formulas (\ref{O1j4}), 
(\ref{Oalfa1}) and (\ref{OLe1}), respectively.

Inserting (\ref{Sigmae1}) into (\ref{lcN02a}),  
the denominator (\ref{vd3}) take the final form:
\begin{equation} \label{vdfinal}
\begin{split}
v_d=(-1)^{\tfrac{m(m+1)}{2}} \, i^{m-m_h} \, \frac{\prod\limits_{j=1}^m Z'(\lambda_j)}{\prod\limits_{j=1}^{m_H} Z'(h_j)}\,
e^{\tfrac{1}{2} \hat{\Sigma}_0},
\end{split}
\end{equation}
where 
\begin{equation} \label{hatSigma0}
\begin{split}
\hat{\Sigma}_0&=-\sum\limits_{j=1}^{m_H} \, \hat{\cal O}_1^{(j)}+
\int\limits_{-\infty}^{\infty} \frac{d \lambda}{2 \pi} \, {\cal O}_-(\lambda-i \, \eta) \, Z'(\lambda-i \, \eta)-\\
&-\sum\limits_{\alpha=\pm} \int\limits_{-\infty}^{\infty} \frac{d \lambda}{2 \pi} {\cal O}_\alpha(\lambda+i \, \alpha \, \eta) \,
Z'(\lambda+i \, \alpha \, \eta) \, {\cal F}_\alpha(\lambda+i \, \alpha \, \eta)
-2 \,m \, \ln \left(1+(-1)^{\delta} e^{i \, Z(+\infty)} \right),
\end{split}
\end{equation}
with $\hat{\cal O}_1^{(j)}, \, {\cal O}_\pm(\lambda)$ given in (\ref{hatO1j}) and (\ref{Oalfa1}), respectively.

\section{Summary} \label{SUMMARYsect}

We close the paper by the summary of the results of the preceding sections. The main purpose of the paper 
was to rewrite the Gaudin-formula \cite{Gaudin0,Gaudin1,Korepin} for the norm of Bethe-wave functions of 
pure hole states into a form, which enables one to carry out and investigate the continuum limit. 
 The Gaudin-formula (\ref{norm1}) is a product of two different type of terms.
The first term denoted by $v_0$ is a double product containing 
the Bethe-roots characterizing the state under consideration, while the 
second term is a nontrivial determinant. Its matrix becomes infinitely large in the continuum limit, thus its 
computation is a nontrivial task in this limit. 

The strategy of the computations was to express all quantities in terms of the counting-function (\ref{countfv}) 
of the model, since it is well-known, that this function encodes the positions of all Bethe-roots of the state 
under consideration and it has a well-defined continuum limit, as well \cite{ddv92},\cite{ddv95}. 

The final result for the product type term $v_0=\tfrac{v_n}{v_d}$ of (\ref{norm1}) can be found in the formulas 
(\ref{lnvnfinal}) and (\ref{vdfinal}). The essence of the final result can be summarized by the following 
formal formula: 
%for $v_0:$
 \begin{equation} \label{v0formal}
\begin{split}
v_0=\frac{\prod\limits_{j=1}^{m_H} Z'(h_j)}{\prod\limits_{j=1}^m Z'(\lambda_j)} \times \left( 
\text{\em "complicated integral expressions of} \, Z(\lambda)\text{\em "} \right),
\end{split}
\end{equation}
where $\lambda_j$s are the Bethe-roots and $h_j$s are the holes.

The result for the determinant part of (\ref{norm1}) can be read off from the formulas 
(\ref{detFi1})-(\ref{FiU}), (\ref{TRrel2}), (\ref{detres1}) and (\ref{detcG1}). 
It becomes proportional to some functional determinants:
\begin{equation} \label{F2}
\begin{split}
\text{det}_{ \!\!\!\! \!\!\!\!{\atop m}}\,\, \, \Phi=(-i)^m \, \frac{\prod\limits_{j=1}^m Z'(\lambda_j)}{\prod\limits_{j=1}^{m_H} Z'(h_j)} \, \text{det}(1+\hat{K}) \,\, 
\text{det}_{ \!\!\!\! \!\!\!\!{\atop m_H}}\!\! (Z'(h_j)\, \delta_{jk}-\bar{\cal G}(h_k,h_j)) \,
\lim\limits_{\eta \to 0^+} \text{det}(1-\hat{B}_\eta),
\end{split}
\end{equation}
where the kernel of $\hat{B}_\eta$ is given in (\ref{Beta}), the function $\bar{\cal G}(\lambda,\lambda')$ 
is defined as the solution of the linear integral equation (\ref{bGl2}) 
and the kernel of $\hat{K}$ is given in (\ref{K1}), such that all functional determinants are considered 
in the space $L^2(-\Lambda^-,\Lambda^+)$ with $\pm \Lambda^\pm$ being the positions of the widest Bethe-roots. 
%or holes. 

From (\ref{v0formal}) and (\ref{F2}) one can recognize, that the product $\tfrac{\prod\limits_{j=1}^m Z'(\lambda_j)}{\prod\limits_{j=1}^{m_H} Z'(h_j)}$ cancels from the final result for the 
norm{\footnote{This was the reason, why we did not transformed these products into integral expressions with the help 
of the summation formulas of section \ref{SUMsect}.}}. Apart from this product, all determinants in (\ref{F2}) can be 
evaluated in the continuum limit. Actually, $\text{det}(1+\hat{K})$ is singular in the continuum limit, but 
using $N$ the number of lattice sites as a regulator, it can be computed in the large $N$ limit by 
evaluating an integral expression (\ref{det1pKc}). The remaining two determinants in (\ref{F2}) have well-defined 
continuum limits. On the one hand, as (\ref{1cGc1}) shows, 
\begin{equation} \label{detHcont}
\begin{split}
\text{det}_{ \!\!\!\! \!\!\!\!{\atop m_H}}\!\! (Z'(h_j)\, \delta_{jk}-\bar{\cal G}(h_k,h_j)) 
\stackrel{N \to \infty}{\to} \text{det}_{ \!\!\!\! \!\!\!\!{\atop m_H}}\,\hat{Q}_{jk}=
\text{det}_{ \!\!\!\! \!\!\!\!{\atop m_H}}\!\partial_{h_j}Z(h_k), 
\end{split}
\end{equation}
which is the determinant of the dressed Gaudin-matrix of the solitons.{\footnote{
For more details see section \ref{Gaudinsect}.
%It is the determinant of the matrix 
%defined from the counting-function analogously to the original Gaudin-matrix (\ref{Phiab}), but 
%now the counting-function is considered as if it was a function of the rapidities of the solitons and not
%of the Bethe-roots
}}.
On the other hand in the continuum limit $\text{det}(1-\hat{B}_\eta) \stackrel{N \to \infty}{\to} \text{det}(1-\hat{B}_{\eta,c}),$ 
such that the kernel of $\hat{B}_{\eta,c}$ is given by (\ref{Betac}).
Though this determinant is a functional one, with the help of the Plemelj-formula (\ref{Plemelj}) 
it can be expanded at large values of the finite volume 
of the continuum quantum field theory (\ref{det1pBplem}). This expansion implies that the infinite volume limit 
of this complicated functional determinant becomes simply 1 (\ref{contegy}). 

To close this summary we would like to emphasize the main message of the paper. The summary of the results of this 
paper show that in the continuum limit the norm square of pure hole states is proportional to two determinants.
One of them is a determinant of an $m_H \times m_H$ matrix such that it corresponds to the Gaudin-matrix of the 
dressed solitons ($m_H$ is the number of solitons). The other determinant is a nontrivial functional determinant 
which can be expanded in the large volume limit, such that its infinite volume limit is equal to 1.

%%%%%%%%%%%%%%%%%%%%%%%%%%%%%%%%%%%%%%%%%%%%%%%%%%%%%%%%%%%%%%%%%%%%%%%%%%%%%%%%%%%%%%%%%%%%%%%%%%%%%%%%%%%%%%%%%%%%%%%%%%%%%
%%%%%%%%%%%%%%%%%%%%%%%%%%%%%%%%%%%%%%%%%%%%%%%

%%%%%%%%%%
%%%%%%
%%%%%%%%%%%
%%%%%%%
%%%%%%%
%%%%%%%%

\vspace{0.5cm}
{\tt Acknowledgments}

\noindent 
The author would like to thank Zolt\'an Bajnok and J\'anos Balog for useful discussions.
This work was supported by the Hungarian National
Science Fund NKFIH  (under K116505) and by an MTA-Lend\"ulet Grant.

\appendix

\section{Some properties of $K(\lambda)$} \label{appA}

This short appendix is devoted to summarize some useful representations, identities 
and properties of the function $K(\lambda)$ defined in (\ref{K1}). Here we repeat its definition:
\begin{equation} \label{K1app}
K(\lambda)=\frac{1}{2 \pi} \frac{\sin(2 \, \gamma)}{\sinh(\lambda-i \, \gamma)\, \sinh(\lambda+i \, \gamma)}.
\end{equation}
It can be represented as a Fourier-integral, as well:
\begin{equation} \label{FK1}
\begin{split}
K(\lambda)=\int\limits_{-\infty}^{\infty} \!\! \frac{d \omega}{2 \pi} \, \tilde{K}(\omega)\, e^{-i \, \omega  \lambda}, 
\end{split}
\end{equation}
where the Fourier-transform takes the form:
\begin{equation} \label{FK}
\begin{split}
\tilde{K}(\omega)=\frac{\sinh\left(\tfrac{\pi \, \omega}{2} (1-\tfrac{2 \gamma}{\pi})\right)}{\sinh(\tfrac{\pi \, \omega}{2} )}=\frac{\sinh\left(\tfrac{\pi \, \omega}{2} \tfrac{p-1}{p+1}\right)}{\sinh(\tfrac{\pi \, \omega}{2} )}.
\end{split}
\end{equation}
In the sequel we will mostly use the $\gamma=\tfrac{\pi}{p+1}$ parameterization for the anisotropy parameter.

The function $K(\lambda)$ is either purely positive or purely negative along the real axis with the property, 
that its absolute value is maximal at the origin:
\begin{equation} \label{Kprop1}
\begin{split}
K(\lambda)=\text{sign}(p-1) \, |K(\lambda)|, \qquad |K(\lambda)|\leq |K(0)|, \qquad \lambda \in {\mathbb R},
\end{split}
\end{equation}
where
\begin{equation} \label{Kzero}
\begin{split}
K(0)=\tfrac{1}{\pi} \tan(\tfrac{p-1}{p+1}\tfrac{\pi}{2}).
\end{split}
\end{equation}
From (\ref{FK}) one can show, that:
\begin{equation} \label{KoK0}
\begin{split}
|\tilde{K}(\omega)|\leq |\tilde{K}(0)|<1, \qquad 
\tilde{K}(\omega)\stackrel{\omega \to \pm \infty}{\sim} e^{-\tfrac{\pi}{p+1} |\omega|}.
\end{split}
\end{equation}
It follows that the integral:
\begin{equation} \label{gp}
\begin{split}
g(p)=\int\limits_{-\infty}^{\infty} \!\! \frac{d \omega}{2 \pi} \frac{|\tilde{K}(\omega)|}{1-|\tilde{K}(\omega)|}
\end{split}
\end{equation}
is convergent for any positive values of $p.$ From the definition (\ref{gp}) it also follows, that:
$g(1)=0,$ thus $g(p) \sim (p-1),$ when $p \sim 1.$

\section{Fourier-basis on the $L^2(-\Lambda^-,\Lambda^+)$ Hilbert-space} \label{appB}

In this short appendix, we review the Fourier-basis on the Hilbert space ${\cal H}=L^2(-\Lambda^-,\Lambda^+)$ and some 
related identities. Let $\varphi, \psi \in {\cal H}$. Their inner product is defined by:
\begin{equation} \label{fipszi}
\begin{split}
\langle \varphi|\psi\rangle=\!\! \int\limits_{-\Lambda^-}^{\Lambda^+} \!\!dx \,  \varphi^*(x) \, \psi(x),
\end{split}
\end{equation}
where $*$ denotes complex conjugation.
A complete, orthonormal basis in ${\cal H}$ can be defined by the Fourier-basis $|\psi_n\rangle_{n \in {\mathbb Z}}$:
\begin{equation} \label{fourier1}
\begin{split}
|\psi_n \rangle \to \psi_n(x)=\frac{e^{i \, p_n \, x}}{\sqrt{\Lambda}}, \qquad p_n=\frac{2 \pi }{\Lambda} \, n, \quad n\in {\mathbb Z},
\end{split}
\end{equation}
where we introduced the notation: $\Lambda=\Lambda^++\Lambda^-.$
This basis is orthonormal and complete:
\begin{equation} \label{foc}
\begin{split}
\langle \psi_n| \psi_m \rangle=\delta_{nm}, \qquad \hat{1}=\sum\limits_{n \in {\mathbb Z}} |\psi_n\rangle \langle \psi_n |.
\end{split}
\end{equation}
Here $\hat{1}$ stands for the unit operator in ${\cal H}.$
The completeness relation in coordinate-space takes the form:
\begin{equation} \label{foc1}
\begin{split}
\delta(x-y)= \sum\limits_{n \in {\mathbb Z}} \psi^*_n(x) \, \psi_n(y)= \tfrac{1}{\Lambda} 
\sum\limits_{n \in {\mathbb Z}} e^{i \, p_n \, (x-y)}, 
\qquad x,y \in (-\Lambda^-,\Lambda^+),
\end{split}
\end{equation}
where $\delta(x)$ denotes the Dirac-delta distribution. 
For the computations detailed in the body of the paper one needs to know 
how the trace of an operator $\hat{O}: {\cal H} \to {\cal H}$ can be written in coordinate representation. 
As a consequence of (\ref{foc1}) the trace in coordinate space can be computed by the formula as follows: 
\begin{equation} \label{trO}
\begin{split}
\text{Tr}\, \hat{O}=\sum\limits_{n \in {\mathbb Z}} \langle \psi_n|\hat{O}|\psi_n\rangle= \!\!\int\limits_{-\Lambda^-}^{\Lambda^+} \!\! dx \,\, O(x,x),
\end{split}
\end{equation}
where $O(x,y)$ is the integral kernel corresponding to the operator $\hat{O}.$

%\begin{equation} \label{}
%\begin{split}
%\end{split}
%\end{equation}

%\begin{eqnarray}
%\end{eqnarray}

%\begin{itemize}\item
%\end{itemize}

%%%%%%%%%%%%%%%%%%%%%%%%%%%%%%%%%%%%%%%%%%%%%%%%%%%%%%%%%%%%%%%%%%%%%%%%%%%%%%%%%%%%%%%%%%%
%%%%%%%%%%%%%%%%%%%%%%%%%%%%%%%%%%%%%%%%%%%%%%%%%%%%%%%%%%%%%%%%%%%%%%%%%%%%%%%%%%%%%%%%%%%
%%%%%%%%%%%%%%%%%%%%%%%%%%%%%%%%%%%%%%%%%%%%%%%%%%%%%%%%%%%%%%%%%%%%%%%%%%%%%%%%%%%%%%%%%%%

\end{document}